\documentclass[aps,twocolumn,groupedaddress,amsmath,amssymb]{revtex4-1}
\usepackage[title,titletoc,toc]{appendix}
\usepackage{graphicx}  
\usepackage{graphics}
\usepackage{dcolumn}   
\usepackage{multirow}
\usepackage{bm}        
\usepackage{verbatim}  
\usepackage{subfigure}
\usepackage{hyperref}
\usepackage{float}
\usepackage{epstopdf}
\usepackage{amssymb}
\usepackage{amsmath}
\hypersetup{
	colorlinks=true,
	linkcolor=blue,
	filecolor=magenta,      
	urlcolor=cyan,
}
\usepackage{booktabs}
\usepackage{xr-hyper}
\usepackage{stackengine}
\usepackage[USenglish]{babel}
\usepackage[normalem]{ulem}

\newcommand{\RNum}[1]{\uppercase\expandafter{\romannumeral #1\relax}}

\newcommand{\CAV}[1]{{\color{black} #1}}
\newcommand{\CLA}[1]{{\color{black} #1}}

\newcommand{\TA}[1]{{\color{black} #1}}

\begin{document}
    
    \title{Vesicles with internal active filaments: self-organized propulsion controls shape, motility, and dynamical response}
    
    \author{Clara Abaurrea-Velasco}
    \affiliation{
	Theoretical Soft Matter and Biophysics, Institute of Complex Systems and Institute for Advanced Simulation, Forschungszentrum J\"ulich, D-52425 J\"ulich, Germany
    }
    \email{c.abaurrea@fz-juelich.de; t.auth@fz-juelich.de; g.gompper@fz-juelich.de}
    
    \author{Thorsten Auth}
    \affiliation{
    Theoretical Soft Matter and Biophysics, Institute of Complex Systems and Institute for Advanced Simulation, Forschungszentrum J\"ulich, D-52425 J\"ulich, Germany
    }
    
    \author{Gerhard Gompper}
    \affiliation{
    Theoretical Soft Matter and Biophysics, Institute of Complex Systems and Institute for Advanced Simulation, Forschungszentrum J\"ulich, D-52425 J\"ulich, Germany
    }

    \date{\today}
    
\begin{abstract}
Self-propulsion and navigation due to the sensing of environmental conditions --- such as durotaxis and chemotaxis --- are remarkable properties of biological cells that cannot be modeled by single-component self-propelled particles.  Therefore, we introduce and study ``flexocytes", deformable vesicles with enclosed attached self-propelled pushing and pulling filaments that align due to steric and membrane-mediated interactions. Using computer simulations in two dimensions, we show that the membrane deforms under the propulsion forces and forms shapes mimicking  motile biological cells, such as keratocytes and neutrophils.  When interacting with walls or with interfaces between different substrates, the internal structure of a flexocyte reorganizes, resulting in a preferred angle of reflection or deflection, respectively.  We predict a correlation between motility patterns, shapes, characteristics of the internal forces, and the response to micropatterned substrates and external stimuli.  We propose that engineered flexocytes with desired mechanosensitive capabilities enable the construction of soft-matter microbots.
\end{abstract}
    
\maketitle
    

\section{Introduction}
In active soft matter, collective behavior of many simple mesoscopic agents can lead to complex structure and dynamics. Examples are self-organization of active granular matter \cite{dauchot_membrane_2017,giomi_bots_2013}, clustering of self-propelled colloidal particles \cite{palacci_living_2013}, vorticity in motility assays of microtubules \cite{sumino_microtubules_2012}, and the dance of topological defects in active nematics \cite{dogic_drops_2012}. Living systems, such as motile cells, cell aggregates, and groups of ants carrying cargo, employ similar principles of active self-organization. In addition, they show sensing capability and can process mechanical stimuli, such as chemotaxis for neutrophils \cite{nuzzi_neutrophil_2007} and mechanosensitivity for epithelial cells \cite{wickert_hierarchy_2016}, which are often essential for biological function. Self-organization of force-generating units is underlying motility and sensing of biological cells \cite{svitikina_myosin_1997} and of cellular aggregates \cite{brochard_aggregates_2018}, and turbulent motion in bacterial colonies \cite{peruani_bacteria_2012}. This raises the question about the minimal complexity required to equip physical model systems with sensitivity. The similarities of the behavior of synthetic and biological systems call for a unified description of the underlying physical mechanisms of self-organization, mechanosensing, and reactivity.

Biological cells are model systems where self-organization of the internal active processes has been studied in detail experimentally \cite{jalal_organization}. Motility is controlled mainly by actin polymerization and actomyosin contractility \cite{mogilner_shape_2009,mogilner_rotating_2017}; it is challenging to study theoretically, because of the large length-scale gap between single filaments and entire cells. Therefore, generic continuum models have been developed to predict cell shape and motility on homogenous substrates \cite{mogilner_shape_2009,ziebert_cell_2012,marenduzzo_cell_2015,kruse_lamellipodium_2006,rappel_cell_2010}, on striped substrates substrates and at interfaces \cite{mizuhara1711,manhart1700,lober1312}, as well as for cell-cell collisions \cite{lober1503}. Alternative continuum models with governing equations derived from the filamentous microstructure have been employed to include details of the cytoskeletal organisation \cite{manhart1700,manhart1507}. Models with explicit filaments have been used to study specific biological processes, such as filopodia formation \cite{fletcher_bundling_2008} and lamellipodial waves \cite{weichsel_actin_2010,shlomovitz_membrane_2007}. 

Many self-organized machineries constructed from various active and passive components inherently include regulation and sensing capability. The internal components react differently to stimuli and the machinery can therefore process external information. The interplay between active and spatial self-organization can be found in biological cells and engineered colloidal systems, as well as in biological or artificial systems on significantly larger length scales. For example, diffusiophoretic Janus colloids form spinning superstructures that can autonomously regulate themselves \cite{aubret1811}, groups of ants collectively carry a large cargo to their nest \cite{gelblum_ant_2015}, and microbots in deformable mobile confinement  \cite{deblais_bots_2018} demonstrate complex responses to environmental cues. 

Here, we introduce self-propelled ``flexocytes", a minimal model system to study the motility of mechanosensitive active vesicles based on self-organization of explicit protrusion and retraction forces. The model consists of active filaments in vesicles pulling or pushing on the membrane, and is suitable for single-agent and multi-agent simulations.
We perform Brownian dynamics simulations for this system in the overdamped regime to model substrate friction and internal noise. We find that the filaments in the flexocytes self-organize to reproduce cellular shapes and motility patterns, giving rise to dynamical phase transitions. Furthermore, we show that explicit pulling forces are sufficient to recover behavior of keratocytes: such flexocytes are reflected at walls and deflected at friction interfaces. Motility in our simulations is subject to noise. Interestingly, additional explicit pushing forces lead to less persistent motion, induce trapping at walls, and reduce the deflection at interfaces. Therefore, we propose ``scattering" at walls and interfaces as an approach to probe the internal dynamics and feedback 
loop of reactive self-propelled particles.

\section{Methods}
We study self-propelled rod-like filaments attached to semiflexible membrane  rings two-dimensional systems using Brownian dynamics simulations. The attachment is either in the direction of the propulsion for pushing filaments or in the opposite direction for pulling filaments \cite{abaurrea_rigid-ring_2017}.

\subsection{FILAMENTS}
The system consists of $N_\textrm{r}$ rod-like filaments with length $L_\textrm{r}$, where each filament consists of $n_\textrm{r}$ beads. The filaments are characterized by their center-of-mass positions $\textbf{r}_{\textrm{r},i}$, their orientation angles $\theta_{\textrm{r},i}$ with respect to the $x$ axis, their center-of-mass velocities $\textbf{v}_{\textrm{r},i}$, and their angular velocities $\pmb{\omega}_{\textrm{r},i}$ \cite{abkenar_collective_2013}. Filament positions and orientations are initialized randomly along the membrane, with the only constraint that filaments and membrane cannot overlap.

Filament-filament and filament-membrane interactions are modeled by the repulsive separation-shifted Lennard-Jones (SSLJ) potential for the bead-bead interaction \cite{abkenar_collective_2013}
\begin{equation}
\label{eqn:SSLJ}
\phi(r) = \begin{cases}
4\epsilon\left[\left(\frac{\sigma^2}{\alpha^2 + r^2}\right)^{6} - \left(\frac{\sigma^2}{\alpha^2 + r^2}\right)^{3}\right] + \phi_0   & r \leq r_{\textrm{cut}}\\
0 & r > r_{\textrm{cut}}
\end{cases}
\textrm{,}
\end{equation}
where $r$ is the distance between two beads, $\alpha$ characterizes the capping of the potential, and $\phi_{0}$ shifts the potential to avoid a discontinuity at $r=r_{\textrm{cut}}$. The length $\alpha=\sqrt{2^{1/3}\sigma^2-r_{\textrm{cut}}^2}$ is calculated by requiring the potential minimum to be at $r_{\textrm{cut}}=2.5 \sigma$. Hence $E_\textrm{r}=\phi(0)-\phi(r_{\textrm{cut}})$ is the potential energy barrier. Once $E_\textrm{r}$ has been set, we obtain $\epsilon=\alpha^{12}E/(\alpha^{12}-4\alpha^6\sigma^6+4\sigma^{12})$. Neighbouring beads overlap by $r_\textrm{cut}$, such that the friction for filament-filament sliding is small and no interlocking occurs \cite{abaurrea_rigid-ring_2017}.

\CLA{The P\'eclet number $\rm{Pe}$ quantifies the strength of the self-propulsion force of the filaments. It is defined as ratio between  propulsion force and  thermal forces acting on the filaments, 
	\begin{equation}
	\textrm{Pe} = \frac{L_{\rm{r}}|\mathbf{F}_{\rm{p}}|}{k_{\rm{B}}T} \, .
	\label{eqn:pe}
	\end{equation}}

\subsection{MEMBRANE}
The membrane ring is characterized by its equilibrium radius $R_\textrm{m}$, bending rigidity $\kappa$, target area $\pi R_{\rm m}^2$, compression modulus $k_{\rm A}$, and perimeter-length compression modulus $k_{\rm S}$.
It is discretized into $n_\textrm{b}$ beads that are separated from each other by a distance $0.2r_{\textrm{cut}}$, such that the membrane is smooth and the friction between the filaments and the membrane is minimal \cite{abaurrea_rigid-ring_2017}.

The membrane beads are connected by harmonic bonds with spring constant $k_S$ and rest length $l_\textrm{m}=2R_\textrm{m}\sin{\left(\pi/n_\textrm{b}\right)}$. The total stretching energy
\begin{equation}
E_S=\frac{k_{S}}{2} \sum_{i=0}^{n_\textrm{b}-1} \left(|\mathbf{R}_i|-l_{\rm m} \right)^2 
\end{equation}
thus controls both bond length and total contour length. Here, $\mathbf{R}_i=\mathbf{r}_{i+1}-\mathbf{r}_i$ is the bond vector from monomer $i$ to monomer $i+1$.
In our simulations, $k_S$ is chosen sufficiently large to prevent changes of the membrane perimeter that are larger than $5\%$.

The bending energy \cite{kierfeld_stretching_2004}
\begin{equation}
E_B=\frac{\kappa}{l_\textrm{m}} \sum_{i=0}^{n_\textrm{b}-1} \left(1 - \frac{\mathbf{R}_{i} \cdot \mathbf{R}_{i+1}}{|\mathbf{R}_{i}| |\mathbf{R}_{i+1}|} \right) \textrm{.}
\label{eq:ebend}
\end{equation}
controls membrane shape and deformations. Here, $\kappa$ is bending rigidity. 

The area-compression energy
\begin{equation}
E_{A} = -\frac{k_{A}}{2} \left(\sum_{i=0}^{n_\textrm{b}-1} A_i - \pi R_\textrm{m}^2 \right)^2 
\end{equation}
controls the area enclosed by the membrane.
Here, $k_A$ is the bulk modulus and $A_\textrm{m}$ is the membrane target area. The area $A_i$ of triangle $i$ is
\CLA{
	\begin{equation}
	A_i=\frac{1}{2} \left| \mathbf{r}_i\times(\mathbf{r}_{i+1}-\mathbf{r}_i) \right|=\frac{1}{2} \left| \mathbf{r}_i\times\mathbf{r}_{i+1} \right| \, \textrm{,}
	\end{equation}
	calculated using the position vectors $\mathbf{r}_i\textrm{ and }\mathbf{r}_{i+1}$ for the membrane beads $i$ and $i+1$, respectively.}

\subsection{FILAMENT-MEMBRANE INTERACTION}
\begin{figure*}
	\centering
	\includegraphics[width=0.9\textwidth]{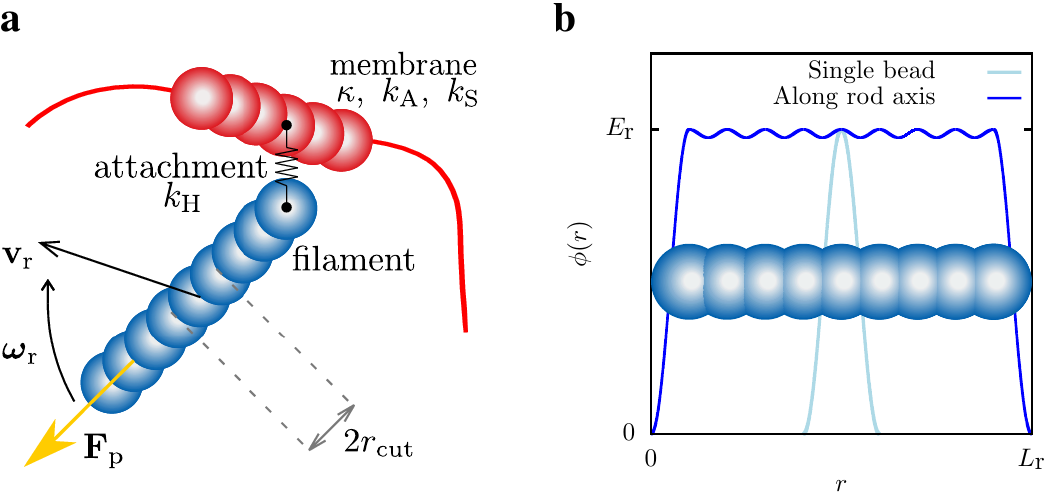} 
	\caption{\CLA{Schematic filament and membrane representation. a) Interaction between a pulling self-propelled filament and a membrane. The first bead of each filament is always attached to the closest bead of the membrane via a harmonic spring. The attachment keeps the filament close to the membrane but allows it to slide. b) Potential profile along a filament. $E_\textrm{r}$ is the energy barrier and $L_\textrm{r}$ the filament length. The light-blue curve represents the potential for a single bead, the dark-blue curve the sum of the contributions for all beads.}}
	\label{fig:rod_vesicle}
\end{figure*}

Filaments and membrane interact sterically via the SSLJ potential with a large energy barrier $E_\textrm{rm}$, such that the filaments cannot exit the membrane ring for all considered propulsion forces.
The filaments are attached to the membrane with their first bead, see Fig.\ref{fig:rod_vesicle}. The attachment is modeled via a harmonic-spring potential
\begin{equation}
\phi_\textrm{att}(r) = 
\begin{cases}
\frac{1}{2}k_\textrm{H}\left(r-b_\textrm{H}\right)^2\textrm{,}   & r \leq r_{\textrm{cut}}\\
\infty\textrm{,} & r > r_{\textrm{cut}}
\end{cases} \textrm{,}
\label{eqn:spring_harm}
\end{equation}
where $k_\textrm{H}$ is the spring constant and $b_\textrm{H}$ the rest length of the spring.
This attachment controls the radial distance between the filaments and the membrane, while still allowing the filaments to slide along the membrane \cite{abaurrea_rigid-ring_2017}.

The direction of the propulsion force with respect to the attachment defines whether an attached filament pushes or pulls. For attached-pushing filaments, the propulsion force points towards the membrane, while for attached-pulling filaments, the propulsion force points away from the membrane.
\CLA{Attached-pulling filaments mutually interact bead-wise via the SSLJ potential defined in Eq.~(\ref{eqn:SSLJ}) with an energy barrier $E_\textrm{r,pull}$, and attached-pushing filaments mutually interact with an energy barrier $E_\textrm{r,push}$. Attached-pulling and attached-pushing filaments do not mutually interact.}

\subsection{SIMULATION TECHNIQUE}
Brownian dynamics simulations are employed for all our systems.
The filament velocity is decomposed into parallel and perpendicular components for the center-of-mass velocity, $\textbf{v}_\textrm{r}=\textbf{v}_{\textrm{r},\parallel}+\textbf{v}_{\textrm{r},\perp}$, and the angular velocity $\pmb{\omega}_{\textrm{r}}$. The filament velocities are given by
\begin{widetext}
\begin{eqnarray}
\mathbf{v}_{\textrm{r}_i\parallel} & = & \frac{1}{\gamma_{\textrm{r}\parallel}}\left(\sum_{j\neq i}^{N_\textrm{r}} \mathbf{F}_{\textrm{r}_{i,j}\parallel}+\mathbf{F}_{\textrm{rm}_i\parallel} + \mathbf{F}_{\textrm{att}_i\parallel} + \xi_{\textrm{r}\parallel}\mathbf{e}_{\parallel} + \mathbf{F}_{\textrm{p}}  \right) \\
\mathbf{v}_{r_i\perp} & = & \frac{1}{\gamma_{\textrm{r}\perp}}\left(\sum_{j\neq i}^{N_\textrm{r}} \mathbf{F}_{\textrm{r}_{i,j}\perp} + \mathbf{F}_{\textrm{rm}_i\perp}+\textbf{F}_{\textrm{att}_i\perp} +\xi_{\textrm{r}\perp}\mathbf{e}_{\perp}\right) \\
\boldsymbol{\omega}_{\textrm{r}_i} & = & \frac{1}{\gamma_{\textrm{r}\theta}}\left(\sum_{j\neq i}^{N_\textrm{r}} \mathbf{M}_{\textrm{r}_{i,j}}+
\mathbf{M}_{\textrm{m}_i}+\textbf{M}_{\textrm{att}_i} +\xi_{\textrm{r}\theta}\textbf{e}_{\theta} \right) \textrm{.}
\end{eqnarray}
\end{widetext}
Here, $\textbf{e}_{\parallel}$ and $\textbf{e}_{\perp}$ are unit vectors parallel and perpendicular to the filament orientation, respectively, and $\textbf{e}_{\theta}$ is a vector oriented normal to the plane of filament motion.
$\textbf{F}_\textrm{p}$ is the propulsion force for each filament. $\textbf{F}_{r_{i,j}}$ and $\textbf{M}_{r_{i,j}}$ are the steric
force and the torque from filament $j$ to filament $i$, and $\textbf{F}_{\textrm{rm}_i}$ and $\textbf{M}_{\textrm{rm}_i}$ are the steric force
and the torque from of the membrane on filament $i$, respectively. $\textbf{F}_{\textrm{att}_i}$ and $\textbf{M}_{\textrm{att}_i}$ are the attachment force and torque from the membrane on filament $i$.

The velocity of a membrane bead $i$ is then
\begin{widetext}
\begin{eqnarray}
\textbf{v}_{\textrm{b}_ix} & = & \frac{1}{\gamma_{\textrm{s}}(x)}\left(\sum_{j=1}^{N_\textrm{r}} \textbf{F}_{\textrm{rb}_{j}x} +\sum_{j=1}^{N_\textrm{r}} \textbf{F}_{\textrm{att}_jx} + \sum_{j=-2}^{2}\textbf{F}_{\textrm{bb}_{i,i-j}x} + \xi_{\textrm{s}}(x)\textbf{e}_{x} + \beta\nabla D_\textrm{s}(x) \right) \nonumber \\ \label{eq:BD_bead1} \\
\textbf{v}_{\textrm{b}_iy} & = & \frac{1}{\gamma_{\textrm{s}}(y)}\left(\sum_{j=1}^{N_\textrm{r}} \textbf{F}_{\textrm{rb}_{j}y} +\sum_{j=1}^{N_\textrm{r}} \textbf{F}_{\textrm{att}_jy} + \sum_{j=-2}^{2}\textbf{F}_{\textrm{bb}_{i,i-j}y} + \xi_{\textrm{s}}(y)\textbf{e}_{y} + \beta\nabla D_\textrm{s}(y) \right) \textrm{,} \nonumber \\ \label{eq:BD_bead2}
\end{eqnarray}
\end{widetext}
where $\textbf{F}_{\textrm{rb}_j}$ is the steric force of filament $j$ on the membrane bead $i$, and $\textbf{F}_{\text{bb}_{i,i-j}}$ represents the stretching, bending, and compression forces of the membrane bead.

The filament friction coefficients in three dimensions are $\gamma_{\textrm{r}\parallel}=\gamma_0L_\textrm{r}$, $\gamma_{\textrm{r}\perp}=2\gamma_{\textrm{r}\parallel}$ and $\gamma_{\textrm{r}\theta}=\gamma_{\textrm{r}\parallel}L_\textrm{r}^2/6$, and $\xi_{\textrm{r}\parallel}$, $\xi_{\textrm{r}\perp}$, and $\xi_{\textrm{r}\theta}$ are the corresponding noise terms, respectively;
$\gamma_\textrm{s}$ is the membrane-substrate friction, and $\xi_{\textrm{s}}$ is the corresponding noise.
All noises are drawn from Gaussian distributions with variances $\sigma_i^2=2k_\textrm{B}T\gamma_i/\delta t$, such that the fluctuation-dissipation theorem is fulfilled, at equilibrium \cite{abaurrea_rigid-ring_2017, loewen_spherocylinders_1994}.

For systems where the friction becomes inhomogeneous the Ito-Stratonovich dilemma has to be taken into account \cite{farago_friction_2014,durang_friction_2015}. Here, we modify the Brownian dynamics equations by adding an extra term containing the gradient of the diffusion coefficient $\beta\nabla D(x,y)$, with $D_\textrm{s}(x,y)=k_\textrm{B}T/\gamma_{\textrm{s}}(x,y)$. The value of $\beta$ distinguishes the Ito $\beta=1$, the Stratonovich $\beta=1/2$, and the anti-Ito $\beta=0$ approach. We employ the Ito approach, which has been applied earlier for active Brownian particles in an environment with an anisotropic friction \cite{bechinger_friction_2011}, where it has also been shown that the numerical results coincide with the particle velocities measured in experiments. If the substrate friction is independent of the position, $\gamma_{\rm s}(x,y)=\gamma_{\rm s}$, the last term of Eqs.~(\ref{eq:BD_bead1}) and (\ref{eq:BD_bead2}) disappears because $\nabla D_\textrm{s}(x,y)=0$.

\subsection{FRICTION INTERFACES}
For flexocytes at friction interfaces, the friction interface is taken to be a smoothed-out step function 
\begin{equation}
\gamma_\textrm{s}(x) = \gamma_{\textrm{sl}} + \frac{({\gamma}_{\textrm{sr}}-\gamma_{\textrm{sl}})}{1+\exp{(-b(x-x_\textrm{int}))}}\textrm{,}
\label{eq:inter}
\end{equation}
where $x_\textrm{int}$ is the center of the interface, ${\gamma}_{\textrm{sl}}$ is the membrane-substrate friction for $x<x_\textrm{int}$, ${\gamma}_{\textrm{sr}}$ is the membrane-substrate friction for $x>x_\textrm{int}$ and $b$ characterizes the interface width. In our simulations, we use $x_\textrm{int}=0R_\textrm{m}$, and $b=100/R_\textrm{m}$, which corresponds to an interface width $w_\textrm{int}\approx0.1R_\textrm{m}$; thus the interface is essentially discontinuous on the scale of the flexocyte.

\subsection{SIMULATION PARAMETERS}
The flexocytes enclose $N_\textrm{r,\textrm{push}}=N_\textrm{r,\textrm{pull}}=16, \, 40$ and $64$ pushing and pulling filaments with P\'eclet numbers $\textrm{Pe}=10, \,25, \,50$ and $100$. Each rod-like filament consists of $n_\textrm{r}=9$ beads. The membrane ring with radius $R_\textrm{m}=2.8L_\textrm{r}$ consists of $n_\textrm{b}=395$ beads. The energy barrier for filament-membrane interaction is $E_\textrm{m}=250 \, k_\textrm{B}T$, which prevents the filaments from escaping from the vesicle. The energy barrier between pulling filaments is $E_\textrm{r,pull}=1\,k_\textrm{B}T$, and between pushing filaments $E_\textrm{r,push}=10\, k_\textrm{B}T$. Pushing and pulling filaments do not interact with each other.

For the membrane, we employ reduced membrane-substrate frictions $\tilde{\gamma}_{\rm s}=\gamma_{\rm s}/\gamma_{0}=1, \,5, \,10$ and $25$, reduced area compression moduli $\tilde{k}_{\rm A}=k_{\rm A}/k_{\rm B}T/R_\textrm{m}^4=1,$ and $100$, a reduced bending rigidity $\tilde{\kappa}=\kappa/(k_\textrm{B}TR_\textrm{m})=2$, and a reduced spring constant $\tilde{k}_{\rm s}=k_{\rm s}/(k_\textrm{B}TR_\textrm{m}^2)=\CLA{1,000}$. The harmonic springs that attach an end bead of a filament to the nearest bead of the membrane has rest length $b_{\rm H}=0.1L_\textrm{r}$ and spring constant $k_{\rm H}=80k_\textrm{B}T/L_\textrm{r}^2$.

\section{Results}
\subsection{FLEXOCYTE MODEL}
Our study is based on a two-dimensional (2D) model system consisting of semiflexible membrane rings, which represent the contact line between the vesicle and the substrate, with $N_\textrm{r}$ attached pushing and pulling filaments,
see Fig.~\ref{fig:snap_shape}. Flexocytes of puller type have only pulling filaments, and flexocytes of mixed type have an equal number of pushing and of pulling filaments.
The membrane ring has an equilibrium radius $R_{\rm m}$ and is subject to a bending elasticity with rigidity $\kappa$, a target area $\pi R_\textrm{m}^2$, an area compressibility with modulus $k_{\rm A}$, 
and a perimeter-length compressibility with modulus $k_{\rm S}$. The filaments are modeled as stiff rods of length $L_{\rm r}$, which are propelled along their
long axis with a force $\mathbf{F}_{\rm P}$. Passive filaments experience the frictions $\gamma_{\textrm{r}\parallel}$ parallel and $\gamma_{\textrm{r}\perp}$ perpendicular to their long axis, and are subject to rotational diffusion with single-rod autocorrelation time $\tau_0$.
Filament length and mutual penetrability determine filament alignment. We use a high penetrability for pulling filaments and a low penetrability for pushing filaments.
For details, see Methods section.

In applications of such a flexocyte model to crawling biological cells on a substrate, the membrane ring 
represents the contact line where the membrane detaches from the substrate, the area the contact interface with the substrate, and the pushing and pulling 
filaments actin polymerization and actomyosin contraction, respectively.
The target area of the contact interface in cells is determined by adhesion, actin polymerization at the lamellipodium, and the elastic energy of membrane and cortical cytoskeleton.
\CLA{Filament attachment to the membrane can be achieved via protein complexes, in which, for example, proteins of the ERM family, formin, and myosin play important roles \cite{bassereau_myosin1_2014,bershadsky_formin_2016,bosk_2011_ezrin}.}
The membrane-substrate friction $\gamma_{\rm s}$ represents the rates of breaking and forming receptor-ligand bonds at rear and front, respectively, when the cell moves. 

In the following, we present our results in terms of 
the reduced membrane-substrate 
friction $\tilde{\gamma}_{\rm s}=\gamma_{\rm s}/\gamma_{0}$ \CLA{per bead}, the reduced area compression 
modulus $\tilde{k}_{\rm A}=k_{\rm A} R_\textrm{m}^4/k_{\rm B}T$, where $k_{\rm B}T$ is the thermal energy with an effective temperature $T$, and
the filament propulsion force characterized by the P\'eclet number $\textrm{Pe}=|\mathbf{F}_\textrm{P}|L_\textrm{r}/k_{\rm B}T$.

\subsection*{SHAPES AND MOTILITY}
\begin{figure*}
	\centering
	\includegraphics[width=0.95\textwidth]{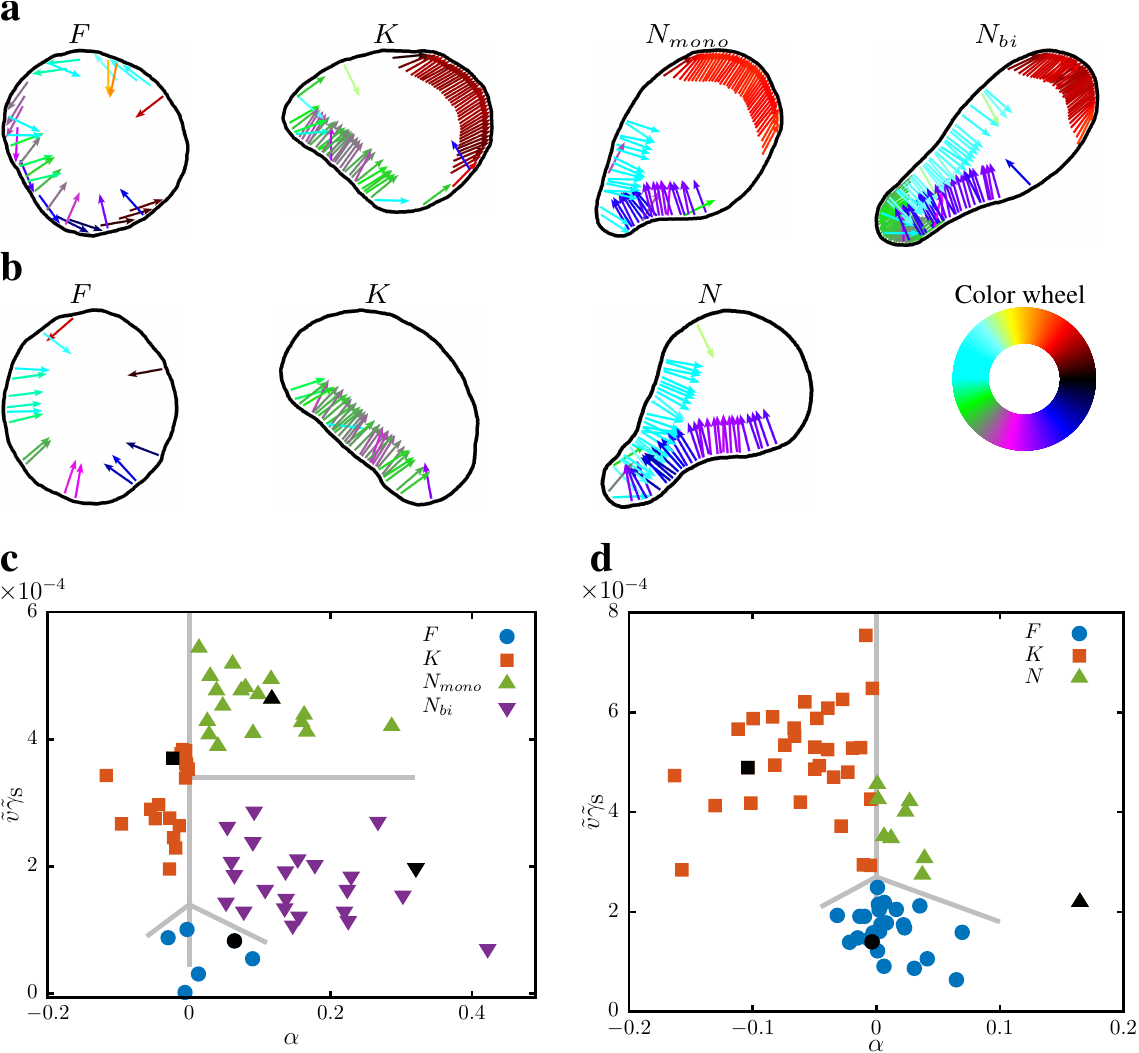} 
	\caption{Shape and motility. \textbf{a)} Flexocytes of mixed type: a $F$-flexocyte with $N_\textrm{r, push}=N_\textrm{r, pull}=16$, $\textrm{Pe}=25$, $\tilde{k}_A=100$, $\tilde{\gamma}_\textrm{s}=5$, a $K$-flexocyte with $N_\textrm{r, push}=N_\textrm{r, pull}=40$, $\textrm{Pe}=50$, $\tilde{k}_A=1$, $\tilde{\gamma}_\textrm{s}=1$, a $N_{mono}$-flexocyte with $N_\textrm{r, push}=N_\textrm{r, pull}=40$, $\textrm{Pe}=50$, $\tilde{k}_A=1$, $\tilde{\gamma}_\textrm{s}=10$, a $N_{bi}$-flexocyte with $N_\textrm{r, push}=N_\textrm{r, pull}=64$, $\textrm{Pe}=100$, $\tilde{k}_A=1$, $\tilde{\gamma}_\textrm{s}=10$. \textbf{b)} Flexocytes with the same parameters as in subfigure (a), but of puller type. \textbf{c)} Scatter plots for the product of flexocyte velocity and membrane-substrate friction $\tilde{v}\tilde{\gamma}_\textrm{s}$, and the signed asphericity $\alpha$. The lines are guides to the eye. Flexocytes of mixed type with $N_\textrm{r}=16$ to 128, $\textrm{Pe}=10$ to 100, $\tilde{k}_A=1$ and 100, and $\tilde{\gamma}_\textrm{s}=1$ to 25. \textbf{d)} Same plot as in subfigure (c) but with flexocytes of puller type. Fluctuating flexocytes (blue circles), keratocytes (orange squares), neutrophils with none or a single cluster of pushing filaments (upward green triangles), and neutrophils with two clusters of pushing filaments (downward purple triangles) occupy different parts of the phase space. The open black symbols correspond to the flexocytes shown in subfigures (a) and (b). See movies M1-M7 in the \CLA{supplementary material}.}
	\label{fig:snap_shape}
\end{figure*}

Flexocytes that are initialized as circular membrane rings with randomly positioned and oriented pushing and pulling filaments assume various stationary shapes, see Fig.~\ref{fig:snap_shape}(a) and (b). For filament propulsion forces that are weak compared with the membrane elastic and friction forces, filaments that pull on the membrane rings point toward the flexocyte centers, while filaments that push circle along the boundaries \cite{abaurrea_rigid-ring_2017}, both leading to nearly stationary quasi-circular, fluctuating ``$F$" shapes. For strong propulsion forces, cluster formation of filaments deforms and propels the flexocytes. For small membrane-substrate frictions motile flexocytes have keratocyte-like ``$K$'' shapes with round apical (front) and flat dorsal (rear) ends \cite{barnhart_adhesion-dependent_2011}. The flat dorsal ends are stabilised by the membrane-mediated alignment of the pulling filaments. For intermediate and large membrane-substrate frictions the dorsal ends assume pointed neutrophil-like ``$N$'' shapes \cite{nuzzi_neutrophil_2007}. The pointed dorsal ends occur for strong membrane-substrate friction where an instability of the distribution of the filament pulling forces induces a strong deformation force on the membrane ring that overcomes its bending stiffness. For mixed systems, we can further distinguish between neutrophile ``$N_{mono}$'' flexocytes with only one cluster (``mono-modal") of pushing filaments at the apical end, and neutrophile ``$N_{bi}$'' flexocytes with two pushing clusters (``bi-modal"), a large (small) cluster at the apical (dorsal) end, which together elongate the shape and slow down the motion. $N_{bi}$-flexocyte-like cells with two lamellipodia have been reported in populations of slow-moving keratocytes \cite{jurado_retrograde_2005}.

We quantify flexocyte motility by the reduced (dimensionless) center-of-mass velocity
\begin{equation}
\tilde{v}=\frac{v\tau_0}{R_\textrm{m}N_\textrm{r}\textrm{Pe}} \, .
\label{eq:vel}
\end{equation}
The velocity $v$ is the ‘instantaneous’ center-of-mass velocity, 
calculated over a time interval $\Delta t/\tau_0=0.02$. It is scaled by $1/(N_\textrm{r}\textrm{Pe})$ that quantifies the maximum propulsion 
of a flexocyte when 
all filaments point in the same direction. The shapes are characterized by the signed asphericity 
\begin{equation}
\alpha=\pm \frac{(\lambda_1-\lambda_2)^2}{(\lambda_1+\lambda_2)^2} \, ,
\label{eq:asph}
\end{equation}
where $\lambda_1$ and $\lambda_2$ are the eigenvalues of the gyration tensor that correspond to the lengths of the long and  short axes, respectively. The asphericity is positive (negative) for flexocytes where the long axis is oriented parallel (perpendicular) to the direction of motion. 

The stationary shapes and motility of flexocytes are dictated by the interplay of the active forces due to the internal degrees of freedom, the conservative forces due to membrane elasticity and area compressibility, and substrate friction. Importantly, we find strong correlations between emergent shape and motility of flexocytes, see Fig.~\ref{fig:snap_shape}(c) and (d). The various types of flexocytes occupy distinct, well-defined regions in a state diagram in which instantaneous total propulsion force $\tilde{v}\tilde{\gamma}_{\rm S}$  and asphericity $\alpha$ are employed as main parameters \footnote{It is important to note that the universal relation between shape and motility is only obtained when the flexocyte velocity multiplied by the substrate friction is employed as relevant observable, not for the flexocyte velocity. The relevant observable is therefore the instantaneous total propulsion force, $F_{act}=\tilde{v}\tilde{\gamma}_{\textrm{s}}$.}. The sign of the asphericity $\alpha$ distinguishes between $K$-flexocytes, and all other shapes. Flexocytes that are elongated parallel to the direction of motion strongly vary in their motility. Not surprisingly, $F$-flexocytes are the slowest, because they have no polarity, and therefore mainly fluctuate without directed translational motion. $N$-flexocytes are polar, and indeed show a pronounced velocity.  For mixed systems, $N_{mono}$-flexocytes are slower than $K$-flexocytes, but are found for larger values of $\tilde{v}\tilde{\gamma}_{\textrm{s}}$ because they are stable at larger substrate frictions. $N_{bi}$-flexocytes are found at smaller values of $\tilde{v}\tilde{\gamma}_{\textrm{s}}$ because their motility is hindered by the cluster of pushing filaments at their dorsal end. For $N$-flexocytes of puller type, not all filaments point in the direction of motion, which leads to reduced velocities compared with $K$-flexocytes.

\begin{figure*}
	\centering
	\includegraphics[width=0.95\textwidth]{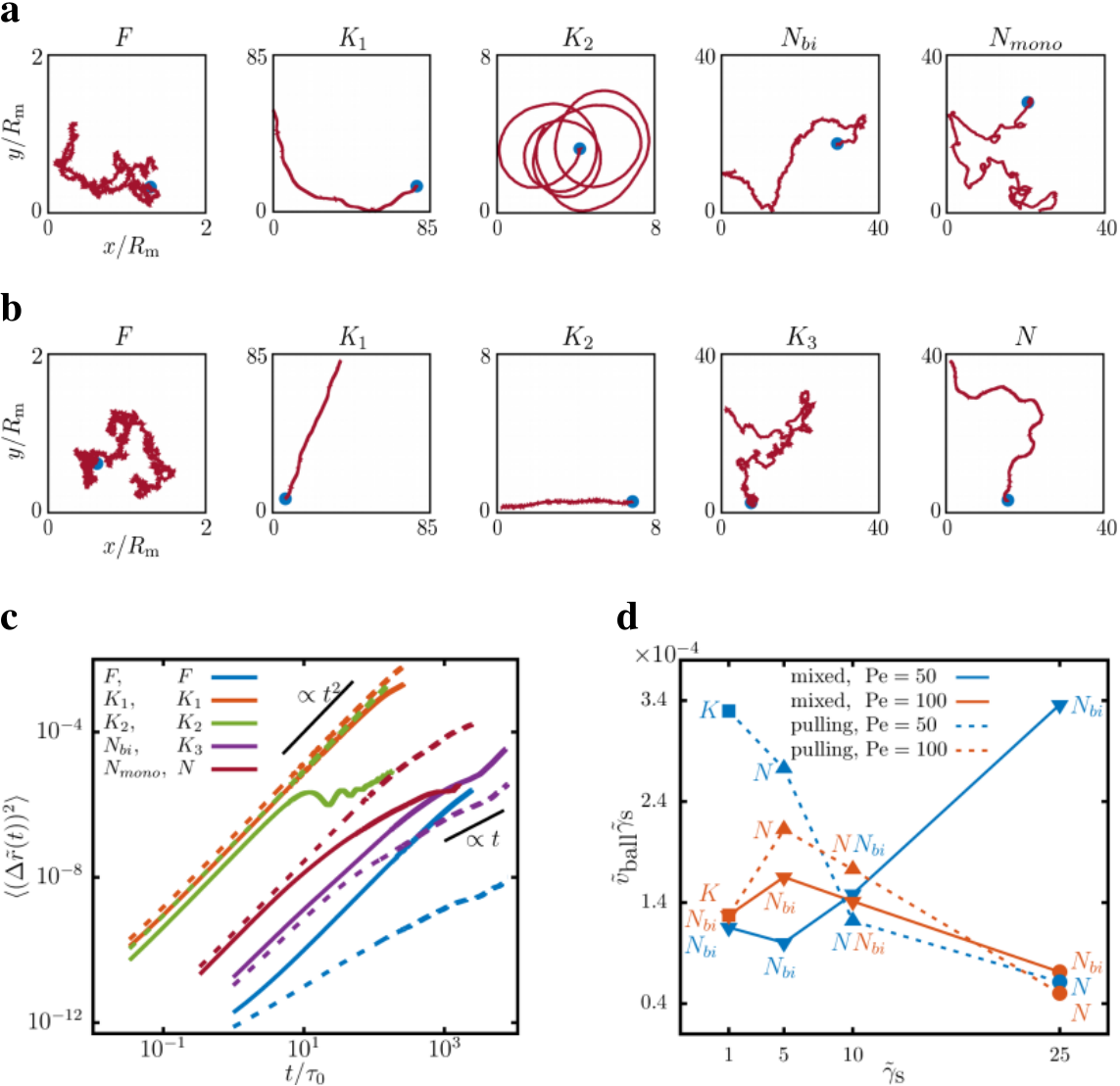} 
	\caption{Trajectories and mean-squared displacements MSDs. 
		\textbf{a)} Center-of-mass trajectories of flexocytes \CLA{simulated for $1045\, \tau_0$}; the starting point is marked by a blue dot. The upper row shows flexocytes of mixed type: $F$-flexocyte with $N_\textrm{r,push}=N_\textrm{r,pull}=64$, $\textrm{Pe}=10$, $\tilde{k}_A=1$, $\tilde{\gamma}_\textrm{s}=25$, 
		a $K$-flexocyte ($K_1$) with $N_\textrm{r,push}=N_\textrm{r,pull}=40$, $\textrm{Pe}=50$, $\tilde{k}_A=1$, $\tilde{\gamma}_\textrm{s}=1$, 
		a $K$-flexocyte ($K_2$) with $N_\textrm{r,push}=N_\textrm{r,pull}=64$, $\textrm{Pe}=25$, $\tilde{k}_A=1$, $\tilde{\gamma}_\textrm{s}=1$, 
		a $N_{bi}$-flexocyte with $N_\textrm{r,push}=N_\textrm{r,pull}=64$, $\textrm{Pe}=100$, $\tilde{k}_A=1$, $\tilde{\gamma}_\textrm{s}=25$, 
		and $N_a$-flexocyte with $N_\textrm{r,push}=N_\textrm{r,pull}=64$, $\textrm{Pe}=100$, $\tilde{k}_A=1$, $\tilde{\gamma}_\textrm{s}=10$. 
		\textbf{b)} Center-of-mass trajectories for the same parameters as the systems in subfigure (a), but for flexocytes of puller type. 
		\textbf{c)} Center-of-mass MSDs of flexocytes shown in subfigures (a) and (b). Solid lines represent flexocytes of mixed type, 
		and dashed lines represent flexocytes of puller type. The left column of the legend refers to flexocytes of mixed type, and the right column refers to flexocytes of puller type. The black lines are guides to the eye showing ballistic $\textrm{MSD}\propto t^2$ 
		and random walk-like $\textrm{MSD}\propto t$ relations. 
		\textbf{d)} Reduced velocities $\tilde{v}_\textrm{ball}=v_\textrm{ball}\tau_0/(R_\textrm{m}N_\textrm{r}\textrm{Pe})$, obtained from the 
		ballistic regime, for different substrate frictions for flexocytes of mixed type $N_\textrm{r, push}=N_\textrm{r, pull}=64$ and 
		of puller type $N_\textrm{r, pull}=64$, $\textrm{Pe}=50$ and 100, and $\tilde{k}_A=1$. The flexocyte shapes are indicated: 
		$F$-flexocytes (circles), $K$-flexocytes (squares), $N$ and $N_{mono}$-flexocytes (upward triangles), and $N_{bi}$ 
		(downward triangles).}
	\label{fig:msd_traj}
\end{figure*}

To study the motility of flexocytes in more detail, we analyze center-of-mass trajectories of flexocytes of mixed type and of puller type for the same parameter values, see Fig.~\ref{fig:msd_traj}(a) and (b), respectively. We characterize the motility by the reduced mean squared displacement
\begin{equation}
\left< \left( \Delta \tilde{r} (t) \right)^2 \right> =\frac{\left<(\mathbf{r}(\tau+t)-\mathbf{r}(\tau))^2\right>}
{\left(R_\textrm{m}N_\textrm{r}\textrm{Pe} \right)^2} \, .
\label{eq:msd}
\end{equation}
As for other self-propelled agents that experience noise, such as active Brownian particles, we expect to find passive diffusive motion at short times, active ballistic motion at intermediate times, and active diffusive motion at long times \cite{golestanian_propulsion_2005, abaurrea_rigid-ring_2017}. For typical $K$-flexocytes of puller type, shown in Fig.~\ref{fig:msd_traj}(b), upon addition of pushing filaments the trajectories become less persistent or circling, and the shape can even change to $N$-flexocytes. 

\begin{figure*}
	\centering
	\includegraphics[width=0.95\textwidth]{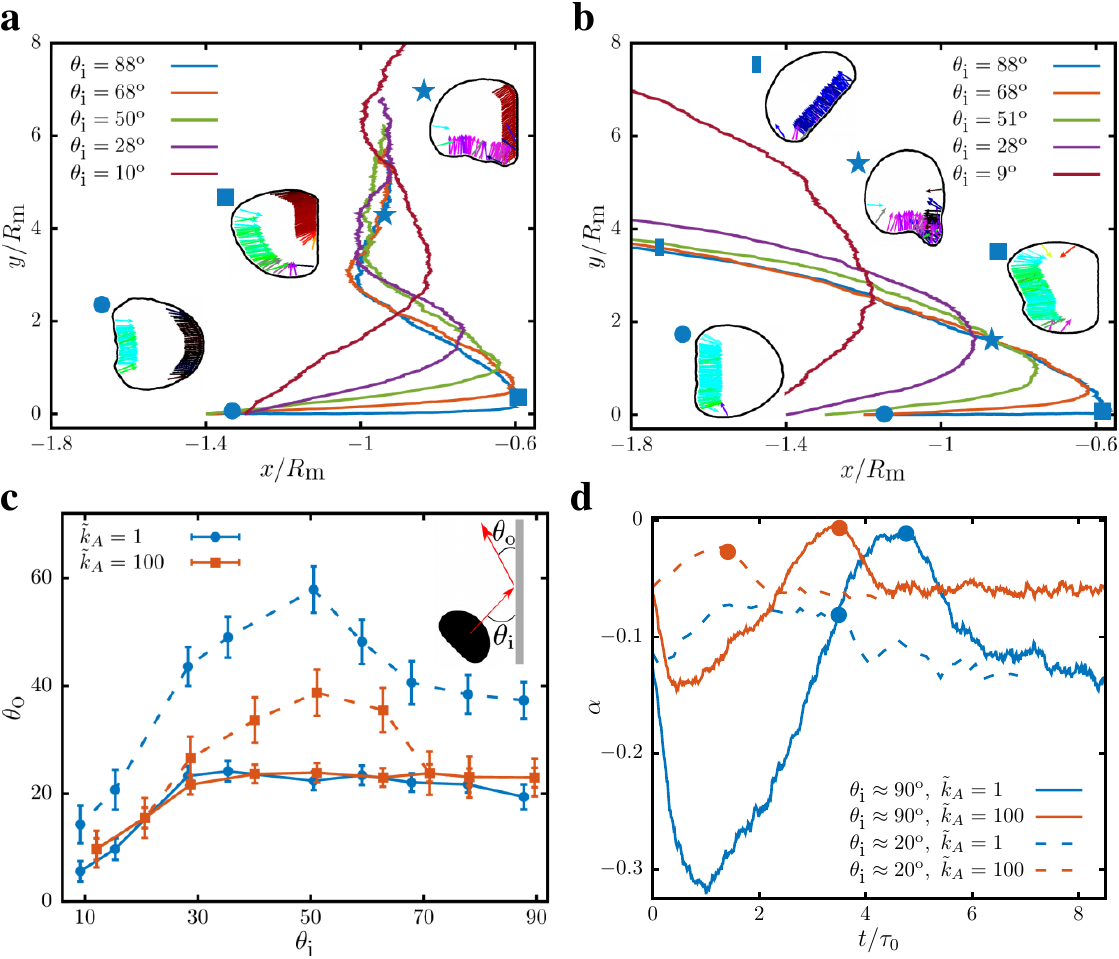} 
	\caption{Flexocytes at walls. 
		\textbf{a)} Center-of-mass trajectories for various angles of incidence $\theta_\textrm{i}$ ($\theta_\textrm{i}$ is measured with respect to the y-axis).
		Flexocyte shapes and orientations for $\theta_\textrm{i}=88\textordmasculine$ are indicated, with symbols refering to the position
		along the trajectory. Flexocytes of mixed type with $N_\textrm{r,push}=N_\textrm{r,pull}=40$, $\textrm{Pe}=50$, $\tilde{k}_A=1$, $\tilde{\gamma}_\textrm{s}=1$. 
		\textbf{b)} Center-of-mass trajectories for various angles of incidence $\theta_\textrm{i}$. Flexocytes of puller type $N_\textrm{r,pull}=64$,
		$\textrm{Pe}=50$, $\tilde{k}_A=1$, and $\tilde{\gamma}_\textrm{s}=1$; flexocyte shapes and orientations for
		$\theta_\textrm{i}=88\textordmasculine$ are indicated.
		\textbf{c)} Angles of reflection versus angle of incidence for flexocytes of puller type $N_\textrm{r,pull}=64$, $\textrm{Pe}=50$ and 
		various values of $\tilde{k}_A$ and $\theta_\textrm{i}$. Solid lines indicate the angle of reflection $\theta_{\textrm{o,near}}$ just after the 
		flexocytes have left the wall, dashed lines indicate the angle of reflection $\theta_{\textrm{o,far}}$ after the flexocytes have relaxed to their 
		new steady-state shape.
		\textbf{d)} Time dependence of the signed asphericity of the flexocytes of puller type $N_\textrm{r,pull}=64$, $\textrm{Pe}=50$ and 
		various values of $\tilde{k}_A$ and $\theta_\textrm{i}$. $t=0$ coincides with the time when the vesicle first comes in contact with the wall. 
		The points indicate the time when the dorsal end leaves the wall. See movies M8-M11 in the SI.}
	\label{fig:traj_wall}
\end{figure*}

In general, $F$-flexocytes have the lowest MSD in the considered range of delay times, followed by $N$-flexocytes, and $K$-flexocytes, 
see Fig.~\ref{fig:msd_traj}(c), consistent with the shape-motility diagrams in Figs.~\ref{fig:snap_shape}(c) and (d). For 
the same parameters, selected flexocytes of mixed type and of puller type show similar propulsion velocities, but 
different persistence of their motion. The stable $K_1$-flexocyte of puller type shows very persistent ballistic motion. 
Additional pushing filaments at the apical end act as a fluctuating steering wheel that destabilises the direction of motion. 
The $K_2$-flexocyte of mixed type even shows circling motion, with periodic oscillations as characteristic signature in the MSD; the superimposed 
linear increase indicates an overall drift. Here, a dynamic instability of the pushing filaments 
breaks the left-right symmetry; the high number of pushing filaments and the motility-induced alignment stabilize the asymmetric shape. The $K_3$-flexocyte, which in parameter space is 
close to $N$-flexocytes, has an unstable dorsal end. Because the pulling filaments may temporarily point in different directions, the persistence of motion is strongly reduced. $N$-flexocytes show less persistent motion than stable $K$-flexocytes, because of  
decreased filament alignment and flexocyte motility. However, in $N_{bi}$-flexocytes of mixed type, pushing-filament
clusters at the apical and the dorsal end stretch the flexocytes and increase their persistence of motion. Thus, the 
$N_{bi}$-flexocyte has a longer ballistic regime than the $N_{mono}$-flexocyte.

Figure~\ref{fig:msd_traj}(d) shows the effective propulsion force $\tilde{v}_{\rm ball} \tilde{\gamma}_{\rm S}$ extracted from the ballistic regime of the MSD, $\langle(\Delta \tilde{r})^2\rangle=\tilde{v}_\textrm{ball}^2t^2$, as function of 
$\tilde{\gamma}_{\rm S}$. 
For a simple self-propelled particle, $\tilde{v}_{\rm ball} \tilde{\gamma}_{\rm S}$ is independent of the friction. The dependence on 
membrane-substrate friction thus reflects the 
different internal organisation of the filaments for the various shapes. 
Here, we briefly discuss the two cases with a pronounced dependence on the friction coefficient, both for $\rm Pe=50$. For flexocytes of puller type,
$K$-flexocytes for small $\tilde{\gamma}_s$ are fast movers, because almost all filaments are aligned with the direction of motion.
They become unstable and transform into slower $N$-flexocytes with increasing membrane-substrate friction.
For flexocytes of mixed type, the emergent immediate total propulsion force of the $N_{\rm bi}$
flexocyte interestingly increases with increasing friction $\tilde{\gamma}_s$. This is related to the cluster formation 
from the random initial state, where a slower deformation of the membrane results in a larger apical/dorsal asymmetry in the 
pushing-filament distribution.

\subsection*{FLEXOCYTES AT WALLS}
\begin{figure*}
	\centering
	\includegraphics[width=0.95\textwidth]{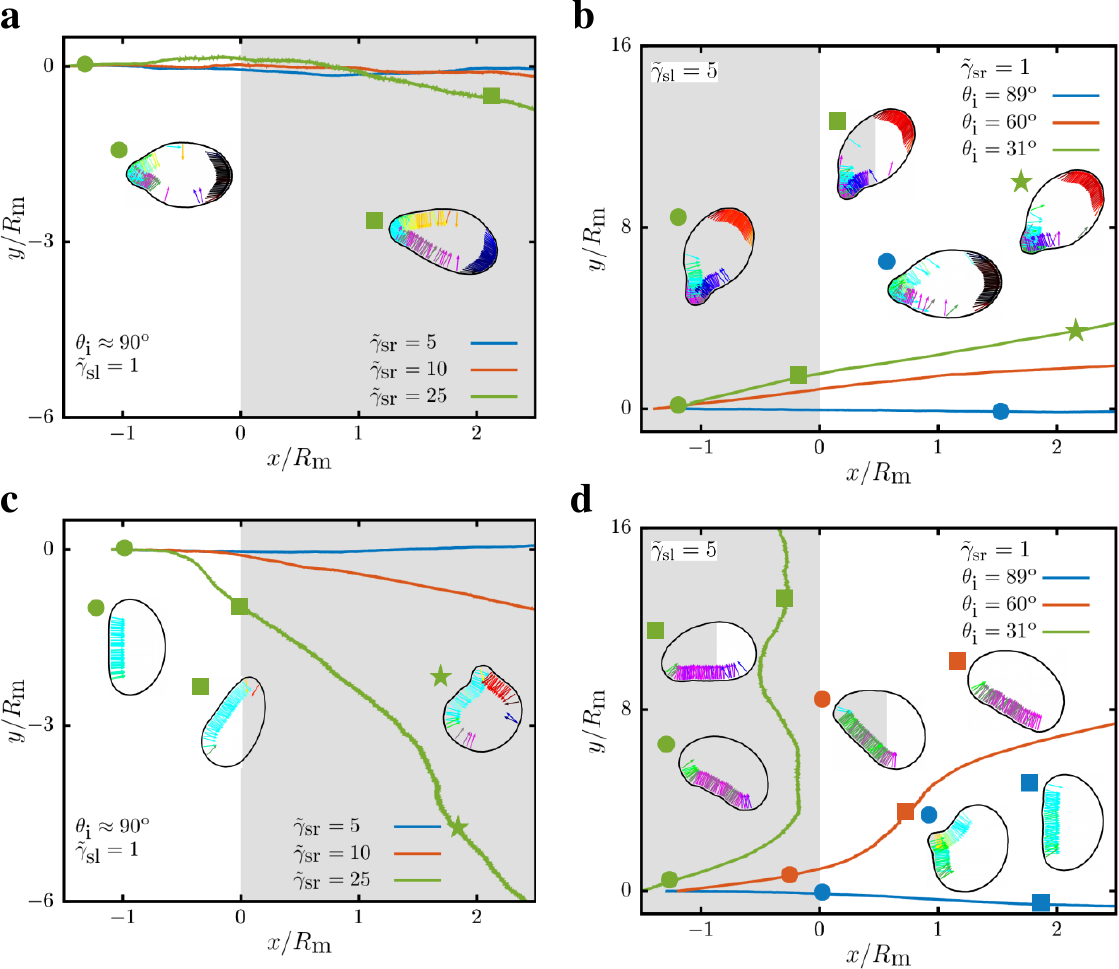}
	\caption{Center-of-mass trajectories of flexocytes at friction interfaces. Flexocytes for various membrane frictions $\tilde{\gamma}_\textrm{s}$ and angles of incidence $\theta_\textrm{i}$. All flexocytes initially move in positive $x$ direction. 
		Selected simulation snapshots indicate instantaneous shapes, with symbols showing corresponding locations on the trajectories.
		\textbf{a)} Flexocytes of mixed type, $N_\textrm{r,push}=N_\textrm{r,pull}=64$, $\textrm{Pe}=100$ and $\tilde{k}_A=100$. Systems with angle of incidence $\theta_\textrm{i}\approx90\textordmasculine$, friction $\tilde{\gamma}_\textrm{s,l}=1$ for $x\leq0$ and various frictions $\tilde{\gamma}_\textrm{s,r}>\tilde{\gamma}_\textrm{s,l}$ for $x>0$. 
		\textbf{b)} Flexocytes with the same parameters as in subfigure (a). Systems with $\tilde{\gamma}_\textrm{s,l}=5$ for $x\leq0$, $\tilde{\gamma}_\textrm{s,r}=1$ for $x>0$, and various $\theta_\textrm{i}$.
		\textbf{c)} Data for the same parameters as in subfigure (a), but for flexocytes of puller type. 
		\textbf{d)} Data for the same parameters as in subfigure (b), but for flexocytes of puller type.
	}
	\label{fig:traj_inter}
\end{figure*}
\begin{figure*}
	\centering
	\includegraphics[width=0.95\textwidth]{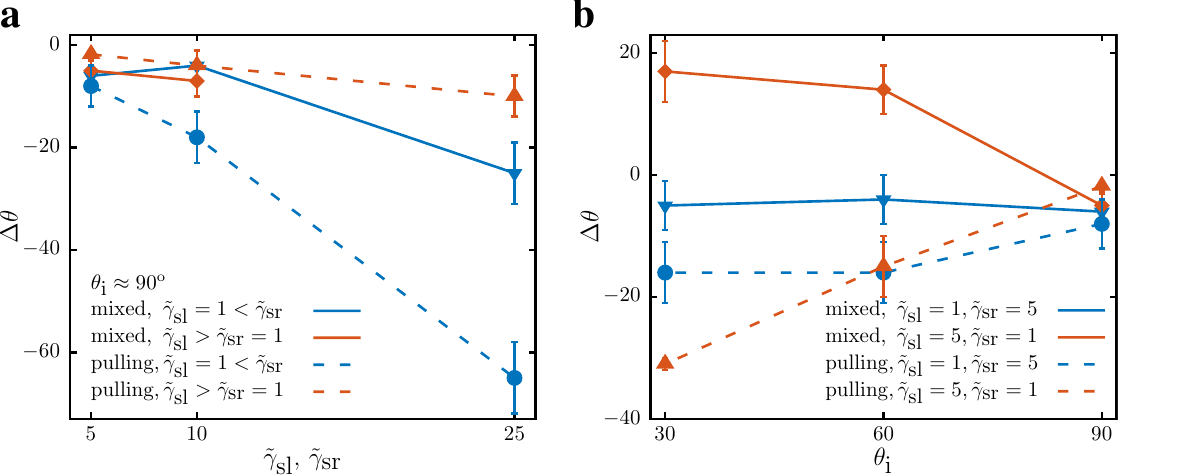}
	\caption{Deflection of trajectories at friction interfaces for various frictions $\tilde{\gamma}_\textrm{s}$ between the membrane and the substrate, and angles of incidence $\theta_\textrm{i}$. 
		\textbf{a)} Deflection $\Delta\theta_\textrm{o}$ versus substrate frictions $\tilde{\gamma}_{\textrm{sl}}$ and $\tilde{\gamma}_{\textrm{sr}}$ for normal angle of incidence $\theta_{\rm i}\approx90\textordmasculine$. Data for flexocytes of mixed type $N_\textrm{r,push}=N_\textrm{r,pull}=64$, $\textrm{Pe}=100$, $\tilde{k}_A=100$, and corresponding data for flexocytes of puller type is shown, averaged over 10 trajectories for each point. For the systems where $\tilde{\gamma}_{\textrm{sl}}=1$, $\tilde{\gamma}_{\textrm{sr}}$ varies and vice-versa. No data exists for the flexocytes of mixed type for $\tilde{\gamma}_\textrm{sl}>\tilde{\gamma}_\textrm{sr}$ and $\tilde{\gamma}_\textrm{sl}=25$. In this case the flexocytes show no broken symmetries which leads to persistent motion and as such they do not cross the friction interface.
		\textbf{b)} Deflection $\Delta\theta_\textrm{o}$ for various angles of incidence $\theta_\textrm{i}$. Analogous to (e), data averaged over 10 trajectores for $\tilde{\gamma}_{\textrm{sl}}=1$ and $\tilde{\gamma}_{\textrm{sr}}=5$, or $\tilde{\gamma}_{\textrm{sl}}=5$ and $\tilde{\gamma}_{\textrm{sr}}=1$, is shown. For positive $\Delta \theta$ the trajectories become more perpendicular to the interface, for negative $\Delta \theta$ more parallel.
	}
	\label{fig:ang_inter}
\end{figure*}

The exposure of flexocytes to external forces and the measurement of their response allows us to probe both their
internal architecture and their dynamic behavior. We first consider the interaction of flexocytes with hard walls and obstacles.
The behavior of the two types of flexocytes is fundamentally different in such an encounter.
Whereas $K$-flexocytes of mixed type get stuck at the walls because of the accumulation of pushing filaments at
the apical end,  see Fig.~\ref{fig:traj_wall}(a), those of puller type are reflected, see Fig.~\ref{fig:traj_wall}(b).
Wall adhesion and reflection are accompanied by major shape changes of the $K$-flexocytes.
Flexocyte deformation thus links the angles of reflection with the flexocyte elastic and active properties.

Directly after contact with a wall, $K$-flexocytes of puller and mixed type elongate along their long axis (parallel to the wall).
The symmetric state is unstable, because small fluctuations lead to a symmetry breaking, which is amplified by the propulsion
forces. For flexocytes of puller type, the shapes subsequently become round, both during reorientation
and shortly after the flexocytes detach from the wall. Finally, the flexocytes recover their stationary shapes.
To characterize this evolution, we determine the reflection angle $\theta_{\textrm{o,near}}$ just after detachment from the wall
and the reflection angle $\theta_{\textrm{o,far}}$ after complete shape relaxation, see Fig.~\ref{fig:traj_wall}(c), and the
time evolution of the asphericity $\alpha$, see Fig.~\ref{fig:traj_wall}(d). All trajectories are more tangential to the wall directly after
wall detachment compared with the angle of incidence, {\em i.e.}, $\theta_{\textrm{o,near}}<\theta_{\textrm{i}}$. 
Furthermore, for $\theta_{\textrm{i}}\geq30\textordmasculine$, we find $\theta_{\textrm{o,near}}\approx 20\textordmasculine$ 
independent of the angle of incidence.
The reflection angles $\theta_{\textrm{o,far}}$ after shape relaxation are either comparable to or larger than 
$\theta_{\textrm{o,near}}$.
For almost tangential impact with $\theta_{\rm i}=20\textordmasculine$, the flexocytes do not deform much, as shown by a nearly constant asphericity, and are almost specularly reflected.

Wall absoption and reflection depend on the flexocyte elastic properties,
see Fig.~\ref{fig:traj_wall}(c) and (d). 
Whereas the qualitative behavior 
remains unchanged for varying the compression modulus $\tilde{k}_A$, both 
deflection angle and asphericity change quantitatively. The far-field angle $\theta_{\textrm{o,far}}$ decreases 
with increasing $\tilde{k}_A$, while the near-field angle $\theta_{\textrm{o,far}}$ is hardly affected and nearly 
independent of the angle of incidence for $\theta_{\textrm{i}}\geq30\textordmasculine$. 
This implies that 
because the shapes become more circular, there is less internal reorganization after detachment from the wall. 
Correspondingly, the shape deformation decreases with increasing $\tilde{k}_A$, as signalled by the reduced variation of the asphericity.
Although the initial flattening for normal impact mimics the elastic deformation of a bouncing ball, the microscopic mechanism is very different because the flexocytes lack inertia; they deform due to the persistence of the pulling-filament motion and their deformability.

\subsection*{FLEXOCYTES AT INTERFACES}
Another interesting approach to study the response to external perturbations is the motility on patterned substrates.
We focus here on friction interfaces, such that the flexocytes move from an area of low friction to one of high
friction, or vice versa.

Figure~\ref{fig:traj_inter} shows center-of-mass trajectories of flexocytes at such friction interfaces. The flexocytes are
initialized in their stationary shapes with their directions of motion oriented toward the interface. Trajectories for
$N$-flexocytes of mixed type are only slightly deflected,
see Figs.~\ref{fig:traj_inter}(a) and (b). Here, the pushing filaments at the apical end prevent major shape deformations and
therefore stabilize the direction of motion.
Trajectories for $K$-flexocytes of puller type can be strongly deflected, see Figs.~\ref{fig:traj_inter}(c) and (d),
because the delicate balance of pulling and friction forces is disturbed at the interface, which leads to strong deformations of the apical ends.
This is demonstrated also in Fig.~\ref{fig:ang_inter}(a), which quantifies deflection as a function of friction jump at the interface.
For almost normal angles of incidence, $\theta_{\textrm{i}}\approx90\textordmasculine$,
flexocytes of mixed type, stabilizing pushing filaments can reduce the deflection angle by a factor $1/3$ compared to flexocytes of puller type
\footnote{Positive values of $\Delta\theta$ indicate a deflection towards the $x$-axis and negative values indicate a deflection towards the $y$-axis.}.

It is important to note that shape changes and trajectories at friction interfaces break time-reversal symmetry, {\em i.e.},
the behavior from high to low friction is very different than in the opposite direction.
This is demonstrated in Fig.~\ref{fig:ang_inter}(a) for initial $K$-flexocytes; for normal incidence and increasing friction the 
trajectories can be strongly deflected, while for decreasing friction they remain nearly perpendicular to the interface.
However, trajectories can also be strongly deflected for decreasing friction, in particular for small angles of incidence, 
see Fig.~\ref{fig:ang_inter}(b).
For $\theta_{\textrm{i}} \lesssim 30\textordmasculine$, after impact the flexocyte attains a state of motion along the interface with 
parts of its membrane in both the large- and small-friction regions. Because the membrane in the small-friction region moves faster, the flexocyte 
settles in a persistent tank-treading motion, which stabilizes the motion along the interfaces and traps it there, see Fig.~\ref{fig:traj_inter}(d).
Interestingly, for flexocytes of puller type friction interfaces in both directions of impact lead to a deflection toward motion parallel to the interface.
The dynamics of transient and persistent shape changes is discussed in the Supplementary Information.

\section*{Discussion}
Responsive active particles can detect and process mechanical information of their environment and show complex responses when they interact with obstacles.
The dynamical phases of our active vesicles with internal self-propelled filaments are dictated by filament steric interactions, membrane deformability, and membrane-substrate friction. 
These flexocytes show feedback between their movement and shape, and are stimuli-sentive to their environment. This is reminiscent of 
motile biological cells, where actin polymerization, contractile actomyosin structures, and membrane elasticity determine both cell 
shape and motility \cite{tojkander_actin_2012,hotulainen_fiber_2006}. 
We show that flexocytes, with pulling filaments at the dorsal end only, are sufficient to mimic shapes and motility of biological cells. 
This agrees well with the importance of pulling forces \CAV{at the back of} keratocytes that break symmetry spontaneously with myosin-driven 
actin flow preceding rear retraction \cite{barnhart_balance_2015}. 
Furthermore, for flexocytes of mixed type, the directional fluctuations of the additional pushing filaments at the apical end can destabilise the flexocyte motion and lead to less ballistic motility compared with flexocytes of puller type. The pushing filaments can even cause circling motion, as has been observed for keratocytes \cite{mogilner_rotating_2017}. In {\it C. elegans} sperm cells, the alignment of the pushing filaments in the lamellipodium is dictated by the membrane tension; their persistence of the motion of the cells increases with increased filament alignment at higher tension \cite{batchelder_tension_2011}.

\TA{Generic models of multi-component active agents elucidate the common generic mechanisms and design principles of responsive biological and engineered active systems. The spontaneous symmetry breaking and self-organisation of the self-propelled filaments is key to the dynamics in our model.}
Micropatterned structures can be used to further characterize shape relaxation, motility, and internal feedback. 
We predict flexocytes of puller type to be reflected by walls with a preferred reflection angle. This finding is very different from 
steady-state accumulation of active Brownian particles at walls \cite{elgeti_surface_2013,brady_pressure_2014}, and it can also not be expected for filament-based cell-motility models that require a manually added chemoattractant gradient or intracellular processes to drive motility \cite{manhart1700,manhart1507}. It reflects the internal reorganisation 
of the filaments and reproduces the behavior of keratocytes at interfaces between adhesive and passivated, microgrooved interfaces 
\cite{miyoshi_characteristics_2012,miyoshi_control_2010}. 
Flexocytes at friction interfaces experience transient and stationary shape changes depending on their internal architecture. The deformation and shape relaxation processes lead to deflection of their trajectories. Our observations for flexocytes of puller type qualitatively reproduce the changes in shape and motility of keratocytes at interfaces between substrates with various adhesion strengths \cite{barnhart_adhesion-dependent_2011}. The simulation results therefore predict an internal force distribution based on observations of cells at interfaces between substrates.

\TA{Unlike most continuum models for biological cells that do not consider noise, flexocytes allow the study of correlations between steady-state shapes, persistence of the trajectories, and force distributions on the membrane. Thereby the simulations provide predictions for cell shape-persistence of trajectory relationships that can be tested in vitro for various cell types. Furthermore, flexocytes predict--bulk or cortex-mediated--cytoskeletal forces on the contact line for various classes of cell shapes. Cytoskeletal organisation can thereby be related to biological function: epithelial cells that benefit from fast movements for wound healing, such as keratocytes, are pulled at the back where all forces point in the direction of movement, whereas cells that follow chemical signals, e.g.\ neutrophils that chase bacteria, are pushed and steered at the front to minimise deflections of the trajectories by mechanical changes of the substrate.}

Our minimal model predicts how sensing capabilities can be realized in engineered active-matter composite agents and in soft microbots. Furthermore, it facilitates studies 
of the active properties of cells, such as an activity-induced tension and active fluctuations of cells \cite{lieber_tension_2013,betz_fluctuations_2016,monzel_fluctuations_2016} and vesicles \cite{dileonardo_vesicle_2016,loubet1203}. 
Future work \TA{in the area of cell motility may include more complex micropatterned substrates \cite{bruckner1903},} many-flexocyte simulations to study collective behavior ranging from the circling observed in small keratinocyte colonies \cite{nanba_rotation_2015} to wound healing \cite{wickert_hierarchy_2016}, and a coupling of mechanics and activity to biochemical reactions \cite{banerjee1506,gross1903}.
In some biological cells, such as neutrophils, cytoskeletal filaments that exert strongly localized forces on lipid-bilayer membranes lead to the formation of sheet-like (lamellipodia) and tube-like (filopodia) protrusions \cite{fritz-laylin17}, whereas in other cells, such as keratocytes, filopodia are absent. Simulations of vesicles in three-dimensions, with two-dimensional membranes, have been employed to study the interplay of cytoskeletal forces, protrusions, and cell motility.
In soft robotics, for systems with active driving forces versatility originates from materials that adjust to shapes and surface properties of the objects that they interact with \cite{whitesides1803}. Future work in the area of robotics may be based on active agents in confinement, such as active colloidal particles in emulsion droplets that have been proposed as model system for cytoplasm on the microscale \cite{horowitz1806}.

\section*{Acknowledgements}
C.A.V. acknowledges support by the International Helmholtz Research School of Biophysics and Soft Matter (IHRS BioSoft). CPU time allowance from the J{\"u}lich Supercomputing Centre (JSC) is gratefully acknowledged. We thank J.\ Plastino (Paris) and A.\ Bershadsky (Singapore/Rehovot) for helpful discussions.
\vspace{3ex}

\bibliography{biblio}

\begin{thebibliography}{63}%
\makeatletter
\providecommand \@ifxundefined [1]{%
 \@ifx{#1\undefined}
}%
\providecommand \@ifnum [1]{%
 \ifnum #1\expandafter \@firstoftwo
 \else \expandafter \@secondoftwo
 \fi
}%
\providecommand \@ifx [1]{%
 \ifx #1\expandafter \@firstoftwo
 \else \expandafter \@secondoftwo
 \fi
}%
\providecommand \natexlab [1]{#1}%
\providecommand \enquote  [1]{``#1''}%
\providecommand \bibnamefont  [1]{#1}%
\providecommand \bibfnamefont [1]{#1}%
\providecommand \citenamefont [1]{#1}%
\providecommand \href@noop [0]{\@secondoftwo}%
\providecommand \href [0]{\begingroup \@sanitize@url \@href}%
\providecommand \@href[1]{\@@startlink{#1}\@@href}%
\providecommand \@@href[1]{\endgroup#1\@@endlink}%
\providecommand \@sanitize@url [0]{\catcode `\\12\catcode `\$12\catcode
  `\&12\catcode `\#12\catcode `\^12\catcode `\_12\catcode `\%12\relax}%
\providecommand \@@startlink[1]{}%
\providecommand \@@endlink[0]{}%
\providecommand \url  [0]{\begingroup\@sanitize@url \@url }%
\providecommand \@url [1]{\endgroup\@href {#1}{\urlprefix }}%
\providecommand \urlprefix  [0]{URL }%
\providecommand \Eprint [0]{\href }%
\providecommand \doibase [0]{http://dx.doi.org/}%
\providecommand \selectlanguage [0]{\@gobble}%
\providecommand \bibinfo  [0]{\@secondoftwo}%
\providecommand \bibfield  [0]{\@secondoftwo}%
\providecommand \translation [1]{[#1]}%
\providecommand \BibitemOpen [0]{}%
\providecommand \bibitemStop [0]{}%
\providecommand \bibitemNoStop [0]{.\EOS\space}%
\providecommand \EOS [0]{\spacefactor3000\relax}%
\providecommand \BibitemShut  [1]{\csname bibitem#1\endcsname}%
\let\auto@bib@innerbib\@empty
\bibitem [{\citenamefont {Junot}\ \emph {et~al.}(2017)\citenamefont {Junot},
  \citenamefont {Briand}, \citenamefont {Ledesma-Alonso},\ and\ \citenamefont
  {Dauchot}}]{dauchot_membrane_2017}%
  \BibitemOpen
  \bibfield  {author} {\bibinfo {author} {\bibfnamefont {G.}~\bibnamefont
  {Junot}}, \bibinfo {author} {\bibfnamefont {G.}~\bibnamefont {Briand}},
  \bibinfo {author} {\bibfnamefont {R.}~\bibnamefont {Ledesma-Alonso}}, \ and\
  \bibinfo {author} {\bibfnamefont {O.}~\bibnamefont {Dauchot}},\ }\href@noop
  {} {\bibfield  {journal} {\bibinfo  {journal} {Phys. Rev. Lett.}\ }\textbf
  {\bibinfo {volume} {119}},\ \bibinfo {pages} {028002} (\bibinfo {year}
  {2017})}\BibitemShut {NoStop}%
\bibitem [{\citenamefont {Giomi}\ \emph {et~al.}(2013)\citenamefont {Giomi},
  \citenamefont {Hawley-Weld},\ and\ \citenamefont
  {Mahadevan}}]{giomi_bots_2013}%
  \BibitemOpen
  \bibfield  {author} {\bibinfo {author} {\bibfnamefont {L.}~\bibnamefont
  {Giomi}}, \bibinfo {author} {\bibfnamefont {N.}~\bibnamefont {Hawley-Weld}},
  \ and\ \bibinfo {author} {\bibfnamefont {L.}~\bibnamefont {Mahadevan}},\
  }\href@noop {} {\bibfield  {journal} {\bibinfo  {journal} {Proc. Royal Soc.
  A}\ }\textbf {\bibinfo {volume} {469}} (\bibinfo {year} {2013})}\BibitemShut
  {NoStop}%
\bibitem [{\citenamefont {Palacci}\ \emph {et~al.}(2013)\citenamefont
  {Palacci}, \citenamefont {Sacanna}, \citenamefont {Steinberg}, \citenamefont
  {Pine},\ and\ \citenamefont {Chaikin}}]{palacci_living_2013}%
  \BibitemOpen
  \bibfield  {author} {\bibinfo {author} {\bibfnamefont {J.}~\bibnamefont
  {Palacci}}, \bibinfo {author} {\bibfnamefont {S.}~\bibnamefont {Sacanna}},
  \bibinfo {author} {\bibfnamefont {A.}~\bibnamefont {Steinberg}}, \bibinfo
  {author} {\bibfnamefont {D.}~\bibnamefont {Pine}}, \ and\ \bibinfo {author}
  {\bibfnamefont {P.}~\bibnamefont {Chaikin}},\ }\href@noop {} {\bibfield
  {journal} {\bibinfo  {journal} {Science}\ }\textbf {\bibinfo {volume}
  {339}},\ \bibinfo {pages} {936} (\bibinfo {year} {2013})}\BibitemShut
  {NoStop}%
\bibitem [{\citenamefont {Sumino}\ \emph {et~al.}(2012)\citenamefont {Sumino},
  \citenamefont {Nagai}, \citenamefont {Shitaka}, \citenamefont {Tanaka},
  \citenamefont {Yoshikawa}, \citenamefont {Chat\'e},\ and\ \citenamefont
  {Oiwa}}]{sumino_microtubules_2012}%
  \BibitemOpen
  \bibfield  {author} {\bibinfo {author} {\bibfnamefont {Y.}~\bibnamefont
  {Sumino}}, \bibinfo {author} {\bibfnamefont {K.}~\bibnamefont {Nagai}},
  \bibinfo {author} {\bibfnamefont {Y.}~\bibnamefont {Shitaka}}, \bibinfo
  {author} {\bibfnamefont {D.}~\bibnamefont {Tanaka}}, \bibinfo {author}
  {\bibfnamefont {K.}~\bibnamefont {Yoshikawa}}, \bibinfo {author}
  {\bibfnamefont {H.}~\bibnamefont {Chat\'e}}, \ and\ \bibinfo {author}
  {\bibfnamefont {K.}~\bibnamefont {Oiwa}},\ }\href@noop {} {\bibfield
  {journal} {\bibinfo  {journal} {Nature}\ }\textbf {\bibinfo {volume} {483}},\
  \bibinfo {pages} {448} (\bibinfo {year} {2012})}\BibitemShut {NoStop}%
\bibitem [{\citenamefont {Sanchez}\ \emph {et~al.}(2012)\citenamefont
  {Sanchez}, \citenamefont {Chen}, \citenamefont {DeCamp}, \citenamefont
  {Heymann},\ and\ \citenamefont {Dogic}}]{dogic_drops_2012}%
  \BibitemOpen
  \bibfield  {author} {\bibinfo {author} {\bibfnamefont {T.}~\bibnamefont
  {Sanchez}}, \bibinfo {author} {\bibfnamefont {D.}~\bibnamefont {Chen}},
  \bibinfo {author} {\bibfnamefont {S.}~\bibnamefont {DeCamp}}, \bibinfo
  {author} {\bibfnamefont {M.}~\bibnamefont {Heymann}}, \ and\ \bibinfo
  {author} {\bibfnamefont {Z.}~\bibnamefont {Dogic}},\ }\href@noop {}
  {\bibfield  {journal} {\bibinfo  {journal} {Nature}\ }\textbf {\bibinfo
  {volume} {491}},\ \bibinfo {pages} {431} (\bibinfo {year}
  {2012})}\BibitemShut {NoStop}%
\bibitem [{\citenamefont {Nuzzi}\ \emph {et~al.}(2007)\citenamefont {Nuzzi},
  \citenamefont {Lokuta},\ and\ \citenamefont
  {Huttenlocher}}]{nuzzi_neutrophil_2007}%
  \BibitemOpen
  \bibfield  {author} {\bibinfo {author} {\bibfnamefont {P.~A.}\ \bibnamefont
  {Nuzzi}}, \bibinfo {author} {\bibfnamefont {M.~A.}\ \bibnamefont {Lokuta}}, \
  and\ \bibinfo {author} {\bibfnamefont {A.}~\bibnamefont {Huttenlocher}},\
  }in\ \href@noop {} {\emph {\bibinfo {booktitle} {Adhesion Protein
  Protocols}}}\ (\bibinfo  {publisher} {Humana Press},\ \bibinfo {address}
  {Totowa, NJ},\ \bibinfo {year} {2007})\BibitemShut {NoStop}%
\bibitem [{\citenamefont {Wickert}\ \emph {et~al.}(2016)\citenamefont
  {Wickert}, \citenamefont {Pomerenke}, \citenamefont {Mitchell}, \citenamefont
  {Masters},\ and\ \citenamefont {Kreeger}}]{wickert_hierarchy_2016}%
  \BibitemOpen
  \bibfield  {author} {\bibinfo {author} {\bibfnamefont {L.~E.}\ \bibnamefont
  {Wickert}}, \bibinfo {author} {\bibfnamefont {S.}~\bibnamefont {Pomerenke}},
  \bibinfo {author} {\bibfnamefont {I.}~\bibnamefont {Mitchell}}, \bibinfo
  {author} {\bibfnamefont {K.~S.}\ \bibnamefont {Masters}}, \ and\ \bibinfo
  {author} {\bibfnamefont {P.~K.}\ \bibnamefont {Kreeger}},\ }\href@noop {}
  {\bibfield  {journal} {\bibinfo  {journal} {Sci. Rep.}\ }\textbf {\bibinfo
  {volume} {6}},\ \bibinfo {pages} {20139} (\bibinfo {year}
  {2016})}\BibitemShut {NoStop}%
\bibitem [{\citenamefont {Svitkina}\ \emph {et~al.}(1997)\citenamefont
  {Svitkina}, \citenamefont {Verkhovsky}, \citenamefont {McQuade},\ and\
  \citenamefont {Borisy}}]{svitikina_myosin_1997}%
  \BibitemOpen
  \bibfield  {author} {\bibinfo {author} {\bibfnamefont {T.~M.}\ \bibnamefont
  {Svitkina}}, \bibinfo {author} {\bibfnamefont {A.~B.}\ \bibnamefont
  {Verkhovsky}}, \bibinfo {author} {\bibfnamefont {K.~M.}\ \bibnamefont
  {McQuade}}, \ and\ \bibinfo {author} {\bibfnamefont {G.~G.}\ \bibnamefont
  {Borisy}},\ }\href@noop {} {\bibfield  {journal} {\bibinfo  {journal} {J.
  Cell Biol.}\ }\textbf {\bibinfo {volume} {139}},\ \bibinfo {pages} {397}
  (\bibinfo {year} {1997})}\BibitemShut {NoStop}%
\bibitem [{\citenamefont {Beaune}\ \emph {et~al.}(2018)\citenamefont {Beaune},
  \citenamefont {Blanch-Mercader}, \citenamefont {Douezan}, \citenamefont
  {Dumond}, \citenamefont {Gonzalez-Rodriguez}, \citenamefont {Cuvelier},
  \citenamefont {Ondar{\c c}uhu}, \citenamefont {Sens}, \citenamefont {Dufour},
  \citenamefont {Murrell},\ and\ \citenamefont
  {Brochard-Wyart}}]{brochard_aggregates_2018}%
  \BibitemOpen
  \bibfield  {author} {\bibinfo {author} {\bibfnamefont {G.}~\bibnamefont
  {Beaune}}, \bibinfo {author} {\bibfnamefont {C.}~\bibnamefont
  {Blanch-Mercader}}, \bibinfo {author} {\bibfnamefont {S.}~\bibnamefont
  {Douezan}}, \bibinfo {author} {\bibfnamefont {J.}~\bibnamefont {Dumond}},
  \bibinfo {author} {\bibfnamefont {D.}~\bibnamefont {Gonzalez-Rodriguez}},
  \bibinfo {author} {\bibfnamefont {D.}~\bibnamefont {Cuvelier}}, \bibinfo
  {author} {\bibfnamefont {T.}~\bibnamefont {Ondar{\c c}uhu}}, \bibinfo
  {author} {\bibfnamefont {P.}~\bibnamefont {Sens}}, \bibinfo {author}
  {\bibfnamefont {S.}~\bibnamefont {Dufour}}, \bibinfo {author} {\bibfnamefont
  {M.~P.}\ \bibnamefont {Murrell}}, \ and\ \bibinfo {author} {\bibfnamefont
  {F.}~\bibnamefont {Brochard-Wyart}},\ }\href@noop {} {\bibfield  {journal}
  {\bibinfo  {journal} {Prof. Natl. Acad. Sci. U.S.A.}\ }\textbf {\bibinfo
  {volume} {115}},\ \bibinfo {pages} {12926} (\bibinfo {year}
  {2018})}\BibitemShut {NoStop}%
\bibitem [{\citenamefont {Peruani}\ \emph {et~al.}(2012)\citenamefont
  {Peruani}, \citenamefont {Starru\ss{}}, \citenamefont {Jakovljevic},
  \citenamefont {S\o{}gaard-Andersen}, \citenamefont {Deutsch},\ and\
  \citenamefont {B\"ar}}]{peruani_bacteria_2012}%
  \BibitemOpen
  \bibfield  {author} {\bibinfo {author} {\bibfnamefont {F.}~\bibnamefont
  {Peruani}}, \bibinfo {author} {\bibfnamefont {J.}~\bibnamefont
  {Starru\ss{}}}, \bibinfo {author} {\bibfnamefont {V.}~\bibnamefont
  {Jakovljevic}}, \bibinfo {author} {\bibfnamefont {L.}~\bibnamefont
  {S\o{}gaard-Andersen}}, \bibinfo {author} {\bibfnamefont {A.}~\bibnamefont
  {Deutsch}}, \ and\ \bibinfo {author} {\bibfnamefont {M.}~\bibnamefont
  {B\"ar}},\ }\href@noop {} {\bibfield  {journal} {\bibinfo  {journal} {Phys.
  Rev. Lett.}\ }\textbf {\bibinfo {volume} {108}},\ \bibinfo {pages} {098102}
  (\bibinfo {year} {2012})}\BibitemShut {NoStop}%
\bibitem [{\citenamefont {Jalal}\ \emph {et~al.}(2019)\citenamefont {Jalal},
  \citenamefont {Shi}, \citenamefont {Acharya}, \citenamefont {Huang},
  \citenamefont {Viasnoff}, \citenamefont {Bershadsky},\ and\ \citenamefont
  {Tee}}]{jalal_organization}%
  \BibitemOpen
  \bibfield  {author} {\bibinfo {author} {\bibfnamefont {S.}~\bibnamefont
  {Jalal}}, \bibinfo {author} {\bibfnamefont {S.}~\bibnamefont {Shi}}, \bibinfo
  {author} {\bibfnamefont {V.}~\bibnamefont {Acharya}}, \bibinfo {author}
  {\bibfnamefont {R.~Y.-J.}\ \bibnamefont {Huang}}, \bibinfo {author}
  {\bibfnamefont {V.}~\bibnamefont {Viasnoff}}, \bibinfo {author}
  {\bibfnamefont {A.}~\bibnamefont {Bershadsky}}, \ and\ \bibinfo {author}
  {\bibfnamefont {Y.~H.}\ \bibnamefont {Tee}},\ }\href@noop {} {\bibfield
  {journal} {\bibinfo  {journal} {J. Cell Sci.}\ } (\bibinfo {year}
  {2019})}\BibitemShut {NoStop}%
\bibitem [{\citenamefont {Mogilner}\ and\ \citenamefont
  {Keren}(2009)}]{mogilner_shape_2009}%
  \BibitemOpen
  \bibfield  {author} {\bibinfo {author} {\bibfnamefont {A.}~\bibnamefont
  {Mogilner}}\ and\ \bibinfo {author} {\bibfnamefont {K.}~\bibnamefont
  {Keren}},\ }\href@noop {} {\bibfield  {journal} {\bibinfo  {journal} {Curr.
  Biol.}\ }\textbf {\bibinfo {volume} {19}},\ \bibinfo {pages} {R762} (\bibinfo
  {year} {2009})}\BibitemShut {NoStop}%
\bibitem [{\citenamefont {Nickaeen}\ \emph {et~al.}(2017)\citenamefont
  {Nickaeen}, \citenamefont {Novak}, \citenamefont {Pulford}, \citenamefont
  {Rumack}, \citenamefont {Brandon}, \citenamefont {Slepchenko},\ and\
  \citenamefont {Mogilner}}]{mogilner_rotating_2017}%
  \BibitemOpen
  \bibfield  {author} {\bibinfo {author} {\bibfnamefont {M.}~\bibnamefont
  {Nickaeen}}, \bibinfo {author} {\bibfnamefont {I.~L.}\ \bibnamefont {Novak}},
  \bibinfo {author} {\bibfnamefont {S.}~\bibnamefont {Pulford}}, \bibinfo
  {author} {\bibfnamefont {A.}~\bibnamefont {Rumack}}, \bibinfo {author}
  {\bibfnamefont {J.}~\bibnamefont {Brandon}}, \bibinfo {author} {\bibfnamefont
  {B.~M.}\ \bibnamefont {Slepchenko}}, \ and\ \bibinfo {author} {\bibfnamefont
  {A.}~\bibnamefont {Mogilner}},\ }\href@noop {} {\bibfield  {journal}
  {\bibinfo  {journal} {PLoS Comput. Biol.}\ }\textbf {\bibinfo {volume}
  {13}},\ \bibinfo {pages} {1} (\bibinfo {year} {2017})}\BibitemShut {NoStop}%
\bibitem [{\citenamefont {Ziebert}\ \emph {et~al.}(2012)\citenamefont
  {Ziebert}, \citenamefont {Swaminathan},\ and\ \citenamefont
  {Aranson}}]{ziebert_cell_2012}%
  \BibitemOpen
  \bibfield  {author} {\bibinfo {author} {\bibfnamefont {F.}~\bibnamefont
  {Ziebert}}, \bibinfo {author} {\bibfnamefont {S.}~\bibnamefont
  {Swaminathan}}, \ and\ \bibinfo {author} {\bibfnamefont {I.}~\bibnamefont
  {Aranson}},\ }\href@noop {} {\bibfield  {journal} {\bibinfo  {journal} {J.
  Royal Soc. Interface}\ }\textbf {\bibinfo {volume} {9}},\ \bibinfo {pages}
  {1084} (\bibinfo {year} {2012})}\BibitemShut {NoStop}%
\bibitem [{\citenamefont {Tjhung}\ \emph {et~al.}(2015)\citenamefont {Tjhung},
  \citenamefont {Tiribocchi}, \citenamefont {Marenduzzo},\ and\ \citenamefont
  {Cates}}]{marenduzzo_cell_2015}%
  \BibitemOpen
  \bibfield  {author} {\bibinfo {author} {\bibfnamefont {E.}~\bibnamefont
  {Tjhung}}, \bibinfo {author} {\bibfnamefont {A.}~\bibnamefont {Tiribocchi}},
  \bibinfo {author} {\bibfnamefont {D.}~\bibnamefont {Marenduzzo}}, \ and\
  \bibinfo {author} {\bibfnamefont {M.~E.}\ \bibnamefont {Cates}},\ }\href@noop
  {} {\bibfield  {journal} {\bibinfo  {journal} {Nat. Commun.}\ }\textbf
  {\bibinfo {volume} {6}},\ \bibinfo {pages} {5420} (\bibinfo {year}
  {2015})}\BibitemShut {NoStop}%
\bibitem [{\citenamefont {Kruse}\ \emph {et~al.}(2006)\citenamefont {Kruse},
  \citenamefont {Joanny}, \citenamefont {J\"ulicher},\ and\ \citenamefont
  {Prost}}]{kruse_lamellipodium_2006}%
  \BibitemOpen
  \bibfield  {author} {\bibinfo {author} {\bibfnamefont {K.}~\bibnamefont
  {Kruse}}, \bibinfo {author} {\bibfnamefont {J.}~\bibnamefont {Joanny}},
  \bibinfo {author} {\bibfnamefont {F.}~\bibnamefont {J\"ulicher}}, \ and\
  \bibinfo {author} {\bibfnamefont {J.}~\bibnamefont {Prost}},\ }\href@noop {}
  {\bibfield  {journal} {\bibinfo  {journal} {Phys. Biol.}\ }\textbf {\bibinfo
  {volume} {3}},\ \bibinfo {pages} {130} (\bibinfo {year} {2006})}\BibitemShut
  {NoStop}%
\bibitem [{\citenamefont {Shao}\ \emph {et~al.}(2010)\citenamefont {Shao},
  \citenamefont {Rappel},\ and\ \citenamefont {Levine}}]{rappel_cell_2010}%
  \BibitemOpen
  \bibfield  {author} {\bibinfo {author} {\bibfnamefont {D.}~\bibnamefont
  {Shao}}, \bibinfo {author} {\bibfnamefont {W.-J.}\ \bibnamefont {Rappel}}, \
  and\ \bibinfo {author} {\bibfnamefont {H.}~\bibnamefont {Levine}},\
  }\href@noop {} {\bibfield  {journal} {\bibinfo  {journal} {Phys. Rev. Lett.}\
  }\textbf {\bibinfo {volume} {105}},\ \bibinfo {pages} {108104} (\bibinfo
  {year} {2010})}\BibitemShut {NoStop}%
\bibitem [{\citenamefont {Mizuhara}\ \emph {et~al.}(2017)\citenamefont
  {Mizuhara}, \citenamefont {Berlyand},\ and\ \citenamefont
  {Aranson}}]{mizuhara1711}%
  \BibitemOpen
  \bibfield  {author} {\bibinfo {author} {\bibfnamefont {M.~S.}\ \bibnamefont
  {Mizuhara}}, \bibinfo {author} {\bibfnamefont {L.}~\bibnamefont {Berlyand}},
  \ and\ \bibinfo {author} {\bibfnamefont {I.~S.}\ \bibnamefont {Aranson}},\
  }\href@noop {} {\bibfield  {journal} {\bibinfo  {journal} {Phys. Rev. E}\
  }\textbf {\bibinfo {volume} {96}},\ \bibinfo {pages} {052408} (\bibinfo
  {year} {2017})}\BibitemShut {NoStop}%
\bibitem [{\citenamefont {Manhart}\ \emph {et~al.}(2017)\citenamefont
  {Manhart}, \citenamefont {Oelz}, \citenamefont {Schmeiser},\ and\
  \citenamefont {Sfakianakis}}]{manhart1700}%
  \BibitemOpen
  \bibfield  {author} {\bibinfo {author} {\bibfnamefont {A.}~\bibnamefont
  {Manhart}}, \bibinfo {author} {\bibfnamefont {D.}~\bibnamefont {Oelz}},
  \bibinfo {author} {\bibfnamefont {C.}~\bibnamefont {Schmeiser}}, \ and\
  \bibinfo {author} {\bibfnamefont {N.}~\bibnamefont {Sfakianakis}},\ }in\
  \href@noop {} {\emph {\bibinfo {booktitle} {Modeling Cellular Systems}}}\
  (\bibinfo  {publisher} {Springer},\ \bibinfo {year} {2017})\ pp.\ \bibinfo
  {pages} {141--159}\BibitemShut {NoStop}%
\bibitem [{\citenamefont {L{\"o}ber}\ \emph {et~al.}(2014)\citenamefont
  {L{\"o}ber}, \citenamefont {Ziebert},\ and\ \citenamefont
  {Aranson}}]{lober1312}%
  \BibitemOpen
  \bibfield  {author} {\bibinfo {author} {\bibfnamefont {J.}~\bibnamefont
  {L{\"o}ber}}, \bibinfo {author} {\bibfnamefont {F.}~\bibnamefont {Ziebert}},
  \ and\ \bibinfo {author} {\bibfnamefont {I.~S.}\ \bibnamefont {Aranson}},\
  }\href@noop {} {\bibfield  {journal} {\bibinfo  {journal} {Soft Matter}\
  }\textbf {\bibinfo {volume} {10}},\ \bibinfo {pages} {1365} (\bibinfo {year}
  {2014})}\BibitemShut {NoStop}%
\bibitem [{\citenamefont {L{\"o}ber}\ \emph {et~al.}(2015)\citenamefont
  {L{\"o}ber}, \citenamefont {Ziebert},\ and\ \citenamefont
  {Aranson}}]{lober1503}%
  \BibitemOpen
  \bibfield  {author} {\bibinfo {author} {\bibfnamefont {J.}~\bibnamefont
  {L{\"o}ber}}, \bibinfo {author} {\bibfnamefont {F.}~\bibnamefont {Ziebert}},
  \ and\ \bibinfo {author} {\bibfnamefont {I.~S.}\ \bibnamefont {Aranson}},\
  }\href@noop {} {\bibfield  {journal} {\bibinfo  {journal} {Sci. Rep.}\
  }\textbf {\bibinfo {volume} {5}},\ \bibinfo {pages} {9172} (\bibinfo {year}
  {2015})}\BibitemShut {NoStop}%
\bibitem [{\citenamefont {Manhart}\ \emph {et~al.}(2015)\citenamefont
  {Manhart}, \citenamefont {Oelz}, \citenamefont {Schmeiser},\ and\
  \citenamefont {Sfakianakis}}]{manhart1507}%
  \BibitemOpen
  \bibfield  {author} {\bibinfo {author} {\bibfnamefont {A.}~\bibnamefont
  {Manhart}}, \bibinfo {author} {\bibfnamefont {D.}~\bibnamefont {Oelz}},
  \bibinfo {author} {\bibfnamefont {C.}~\bibnamefont {Schmeiser}}, \ and\
  \bibinfo {author} {\bibfnamefont {N.}~\bibnamefont {Sfakianakis}},\
  }\href@noop {} {\bibfield  {journal} {\bibinfo  {journal} {J. Theor. Biol.}\
  }\textbf {\bibinfo {volume} {382}},\ \bibinfo {pages} {244} (\bibinfo {year}
  {2015})}\BibitemShut {NoStop}%
\bibitem [{\citenamefont {Liu}\ \emph {et~al.}(2008)\citenamefont {Liu},
  \citenamefont {Richmond}, \citenamefont {Maibaum}, \citenamefont {Pronk},
  \citenamefont {Geissler},\ and\ \citenamefont
  {Fletcher}}]{fletcher_bundling_2008}%
  \BibitemOpen
  \bibfield  {author} {\bibinfo {author} {\bibfnamefont {A.~P.}\ \bibnamefont
  {Liu}}, \bibinfo {author} {\bibfnamefont {D.~L.}\ \bibnamefont {Richmond}},
  \bibinfo {author} {\bibfnamefont {L.}~\bibnamefont {Maibaum}}, \bibinfo
  {author} {\bibfnamefont {S.}~\bibnamefont {Pronk}}, \bibinfo {author}
  {\bibfnamefont {P.~L.}\ \bibnamefont {Geissler}}, \ and\ \bibinfo {author}
  {\bibfnamefont {D.~A.}\ \bibnamefont {Fletcher}},\ }\href@noop {} {\bibfield
  {journal} {\bibinfo  {journal} {Nat. Phys.}\ }\textbf {\bibinfo {volume}
  {4}},\ \bibinfo {pages} {789} (\bibinfo {year} {2008})}\BibitemShut {NoStop}%
\bibitem [{\citenamefont {Weichsel}\ and\ \citenamefont
  {Schwarz}(2010)}]{weichsel_actin_2010}%
  \BibitemOpen
  \bibfield  {author} {\bibinfo {author} {\bibfnamefont {J.}~\bibnamefont
  {Weichsel}}\ and\ \bibinfo {author} {\bibfnamefont {U.~S.}\ \bibnamefont
  {Schwarz}},\ }\href@noop {} {\bibfield  {journal} {\bibinfo  {journal} {Proc.
  Natl. Acad. Sci. U.S.A.}\ }\textbf {\bibinfo {volume} {107}},\ \bibinfo
  {pages} {6304} (\bibinfo {year} {2010})}\BibitemShut {NoStop}%
\bibitem [{\citenamefont {Shlomovitz}\ and\ \citenamefont
  {Gov}(2007)}]{shlomovitz_membrane_2007}%
  \BibitemOpen
  \bibfield  {author} {\bibinfo {author} {\bibfnamefont {R.}~\bibnamefont
  {Shlomovitz}}\ and\ \bibinfo {author} {\bibfnamefont {N.~S.}\ \bibnamefont
  {Gov}},\ }\href@noop {} {\bibfield  {journal} {\bibinfo  {journal} {Phys.
  Rev. Lett.}\ }\textbf {\bibinfo {volume} {98}},\ \bibinfo {pages} {168103}
  (\bibinfo {year} {2007})}\BibitemShut {NoStop}%
\bibitem [{\citenamefont {Aubret}\ \emph {et~al.}(2018)\citenamefont {Aubret},
  \citenamefont {Youssef}, \citenamefont {Sacanna},\ and\ \citenamefont
  {Palacci}}]{aubret1811}%
  \BibitemOpen
  \bibfield  {author} {\bibinfo {author} {\bibfnamefont {A.}~\bibnamefont
  {Aubret}}, \bibinfo {author} {\bibfnamefont {M.}~\bibnamefont {Youssef}},
  \bibinfo {author} {\bibfnamefont {S.}~\bibnamefont {Sacanna}}, \ and\
  \bibinfo {author} {\bibfnamefont {J.}~\bibnamefont {Palacci}},\ }\href@noop
  {} {\bibfield  {journal} {\bibinfo  {journal} {Nat. Phys.}\ }\textbf
  {\bibinfo {volume} {14}},\ \bibinfo {pages} {1114} (\bibinfo {year}
  {2018})}\BibitemShut {NoStop}%
\bibitem [{\citenamefont {Gelblum}\ \emph {et~al.}(2015)\citenamefont
  {Gelblum}, \citenamefont {Pinkoviezky}, \citenamefont {Fonio}, \citenamefont
  {Ghosh}, \citenamefont {Gov},\ and\ \citenamefont
  {Feinerman}}]{gelblum_ant_2015}%
  \BibitemOpen
  \bibfield  {author} {\bibinfo {author} {\bibfnamefont {A.}~\bibnamefont
  {Gelblum}}, \bibinfo {author} {\bibfnamefont {I.}~\bibnamefont
  {Pinkoviezky}}, \bibinfo {author} {\bibfnamefont {E.}~\bibnamefont {Fonio}},
  \bibinfo {author} {\bibfnamefont {A.}~\bibnamefont {Ghosh}}, \bibinfo
  {author} {\bibfnamefont {N.}~\bibnamefont {Gov}}, \ and\ \bibinfo {author}
  {\bibfnamefont {O.}~\bibnamefont {Feinerman}},\ }\href@noop {} {\bibfield
  {journal} {\bibinfo  {journal} {Nat. Commun.}\ }\textbf {\bibinfo {volume}
  {6}},\ \bibinfo {pages} {7729} (\bibinfo {year} {2015})}\BibitemShut
  {NoStop}%
\bibitem [{\citenamefont {Deblais}\ \emph {et~al.}(2018)\citenamefont
  {Deblais}, \citenamefont {Barois}, \citenamefont {Guerin}, \citenamefont
  {Delville}, \citenamefont {Vaudaine}, \citenamefont {Lintuvuori},
  \citenamefont {Boudet}, \citenamefont {Baret},\ and\ \citenamefont
  {Kellay}}]{deblais_bots_2018}%
  \BibitemOpen
  \bibfield  {author} {\bibinfo {author} {\bibfnamefont {A.}~\bibnamefont
  {Deblais}}, \bibinfo {author} {\bibfnamefont {T.}~\bibnamefont {Barois}},
  \bibinfo {author} {\bibfnamefont {T.}~\bibnamefont {Guerin}}, \bibinfo
  {author} {\bibfnamefont {P.~H.}\ \bibnamefont {Delville}}, \bibinfo {author}
  {\bibfnamefont {R.}~\bibnamefont {Vaudaine}}, \bibinfo {author}
  {\bibfnamefont {J.~S.}\ \bibnamefont {Lintuvuori}}, \bibinfo {author}
  {\bibfnamefont {J.~F.}\ \bibnamefont {Boudet}}, \bibinfo {author}
  {\bibfnamefont {J.~C.}\ \bibnamefont {Baret}}, \ and\ \bibinfo {author}
  {\bibfnamefont {H.}~\bibnamefont {Kellay}},\ }\href@noop {} {\bibfield
  {journal} {\bibinfo  {journal} {Phys. Rev. Lett.}\ }\textbf {\bibinfo
  {volume} {120}},\ \bibinfo {pages} {188002} (\bibinfo {year}
  {2018})}\BibitemShut {NoStop}%
\bibitem [{\citenamefont {Abaurrea~Velasco}\ \emph {et~al.}(2017)\citenamefont
  {Abaurrea~Velasco}, \citenamefont {Dehghani~Ghahnaviyeh}, \citenamefont
  {Nejat~Pishkenari}, \citenamefont {Auth},\ and\ \citenamefont
  {Gompper}}]{abaurrea_rigid-ring_2017}%
  \BibitemOpen
  \bibfield  {author} {\bibinfo {author} {\bibfnamefont {C.}~\bibnamefont
  {Abaurrea~Velasco}}, \bibinfo {author} {\bibfnamefont {S.}~\bibnamefont
  {Dehghani~Ghahnaviyeh}}, \bibinfo {author} {\bibfnamefont {H.}~\bibnamefont
  {Nejat~Pishkenari}}, \bibinfo {author} {\bibfnamefont {T.}~\bibnamefont
  {Auth}}, \ and\ \bibinfo {author} {\bibfnamefont {G.}~\bibnamefont
  {Gompper}},\ }\href@noop {} {\bibfield  {journal} {\bibinfo  {journal} {Soft
  Matter}\ }\textbf {\bibinfo {volume} {13}},\ \bibinfo {pages} {5865}
  (\bibinfo {year} {2017})}\BibitemShut {NoStop}%
\bibitem [{\citenamefont {Abkenar}\ \emph {et~al.}(2013)\citenamefont
  {Abkenar}, \citenamefont {Marx}, \citenamefont {Auth},\ and\ \citenamefont
  {Gompper}}]{abkenar_collective_2013}%
  \BibitemOpen
  \bibfield  {author} {\bibinfo {author} {\bibfnamefont {M.}~\bibnamefont
  {Abkenar}}, \bibinfo {author} {\bibfnamefont {K.}~\bibnamefont {Marx}},
  \bibinfo {author} {\bibfnamefont {T.}~\bibnamefont {Auth}}, \ and\ \bibinfo
  {author} {\bibfnamefont {G.}~\bibnamefont {Gompper}},\ }\href@noop {}
  {\bibfield  {journal} {\bibinfo  {journal} {Phys. Rev. E}\ }\textbf {\bibinfo
  {volume} {88}},\ \bibinfo {pages} {062314} (\bibinfo {year}
  {2013})}\BibitemShut {NoStop}%
\bibitem [{\citenamefont {Kierfeld}\ \emph {et~al.}(2004)\citenamefont
  {Kierfeld}, \citenamefont {Niamploy}, \citenamefont {Sa-yakanit},\ and\
  \citenamefont {Lipowsky}}]{kierfeld_stretching_2004}%
  \BibitemOpen
  \bibfield  {author} {\bibinfo {author} {\bibfnamefont {J.}~\bibnamefont
  {Kierfeld}}, \bibinfo {author} {\bibfnamefont {O.}~\bibnamefont {Niamploy}},
  \bibinfo {author} {\bibfnamefont {V.}~\bibnamefont {Sa-yakanit}}, \ and\
  \bibinfo {author} {\bibfnamefont {R.}~\bibnamefont {Lipowsky}},\ }\href@noop
  {} {\bibfield  {journal} {\bibinfo  {journal} {Eur. Phys. J. E}\ }\textbf
  {\bibinfo {volume} {14}},\ \bibinfo {pages} {17} (\bibinfo {year}
  {2004})}\BibitemShut {NoStop}%
\bibitem [{\citenamefont {L\"owen}(1994)}]{loewen_spherocylinders_1994}%
  \BibitemOpen
  \bibfield  {author} {\bibinfo {author} {\bibfnamefont {H.}~\bibnamefont
  {L\"owen}},\ }\href@noop {} {\bibfield  {journal} {\bibinfo  {journal} {Phys.
  Rev. E}\ }\textbf {\bibinfo {volume} {50}},\ \bibinfo {pages} {1232}
  (\bibinfo {year} {1994})}\BibitemShut {NoStop}%
\bibitem [{\citenamefont {Farago}\ and\ \citenamefont
  {Gr\o{}nbech-Jensen}(2014)}]{farago_friction_2014}%
  \BibitemOpen
  \bibfield  {author} {\bibinfo {author} {\bibfnamefont {O.}~\bibnamefont
  {Farago}}\ and\ \bibinfo {author} {\bibfnamefont {N.}~\bibnamefont
  {Gr\o{}nbech-Jensen}},\ }\href@noop {} {\bibfield  {journal} {\bibinfo
  {journal} {Phys. Rev. E}\ }\textbf {\bibinfo {volume} {89}},\ \bibinfo
  {pages} {013301} (\bibinfo {year} {2014})}\BibitemShut {NoStop}%
\bibitem [{\citenamefont {Durang}\ \emph {et~al.}(2015)\citenamefont {Durang},
  \citenamefont {Kwon},\ and\ \citenamefont {Park}}]{durang_friction_2015}%
  \BibitemOpen
  \bibfield  {author} {\bibinfo {author} {\bibfnamefont {X.}~\bibnamefont
  {Durang}}, \bibinfo {author} {\bibfnamefont {C.}~\bibnamefont {Kwon}}, \ and\
  \bibinfo {author} {\bibfnamefont {H.}~\bibnamefont {Park}},\ }\href@noop {}
  {\bibfield  {journal} {\bibinfo  {journal} {Phys. Rev. E}\ }\textbf {\bibinfo
  {volume} {91}},\ \bibinfo {pages} {062118} (\bibinfo {year}
  {2015})}\BibitemShut {NoStop}%
\bibitem [{\citenamefont {Brettschneider}\ \emph {et~al.}(2011)\citenamefont
  {Brettschneider}, \citenamefont {Volpe}, \citenamefont {Helden},
  \citenamefont {Wehr},\ and\ \citenamefont
  {Bechinger}}]{bechinger_friction_2011}%
  \BibitemOpen
  \bibfield  {author} {\bibinfo {author} {\bibfnamefont {T.}~\bibnamefont
  {Brettschneider}}, \bibinfo {author} {\bibfnamefont {G.}~\bibnamefont
  {Volpe}}, \bibinfo {author} {\bibfnamefont {L.}~\bibnamefont {Helden}},
  \bibinfo {author} {\bibfnamefont {J.}~\bibnamefont {Wehr}}, \ and\ \bibinfo
  {author} {\bibfnamefont {C.}~\bibnamefont {Bechinger}},\ }\href@noop {}
  {\bibfield  {journal} {\bibinfo  {journal} {Phys. Rev. E}\ }\textbf {\bibinfo
  {volume} {83}},\ \bibinfo {pages} {041113} (\bibinfo {year}
  {2011})}\BibitemShut {NoStop}%
\bibitem [{\citenamefont {Yamada}\ \emph {et~al.}(2014)\citenamefont {Yamada},
  \citenamefont {Mamane}, \citenamefont {Lee-Tin-Wah}, \citenamefont
  {Di~Cicco}, \citenamefont {Pr{\'e}vost}, \citenamefont {L{\'e}vy},
  \citenamefont {Joanny}, \citenamefont {Coudrier},\ and\ \citenamefont
  {Bassereau}}]{bassereau_myosin1_2014}%
  \BibitemOpen
  \bibfield  {author} {\bibinfo {author} {\bibfnamefont {A.}~\bibnamefont
  {Yamada}}, \bibinfo {author} {\bibfnamefont {A.}~\bibnamefont {Mamane}},
  \bibinfo {author} {\bibfnamefont {J.}~\bibnamefont {Lee-Tin-Wah}}, \bibinfo
  {author} {\bibfnamefont {A.}~\bibnamefont {Di~Cicco}}, \bibinfo {author}
  {\bibfnamefont {C.}~\bibnamefont {Pr{\'e}vost}}, \bibinfo {author}
  {\bibfnamefont {D.}~\bibnamefont {L{\'e}vy}}, \bibinfo {author}
  {\bibfnamefont {J.-F.}\ \bibnamefont {Joanny}}, \bibinfo {author}
  {\bibfnamefont {E.}~\bibnamefont {Coudrier}}, \ and\ \bibinfo {author}
  {\bibfnamefont {P.}~\bibnamefont {Bassereau}},\ }\href@noop {} {\bibfield
  {journal} {\bibinfo  {journal} {Nat. Commun.}\ }\textbf {\bibinfo {volume}
  {5}},\ \bibinfo {pages} {3624} (\bibinfo {year} {2014})}\BibitemShut
  {NoStop}%
\bibitem [{\citenamefont {Luo}\ \emph {et~al.}(2016)\citenamefont {Luo},
  \citenamefont {Lieu}, \citenamefont {Manser}, \citenamefont {Bershadsky},\
  and\ \citenamefont {Sheetz}}]{bershadsky_formin_2016}%
  \BibitemOpen
  \bibfield  {author} {\bibinfo {author} {\bibfnamefont {W.}~\bibnamefont
  {Luo}}, \bibinfo {author} {\bibfnamefont {Z.~Z.}\ \bibnamefont {Lieu}},
  \bibinfo {author} {\bibfnamefont {E.}~\bibnamefont {Manser}}, \bibinfo
  {author} {\bibfnamefont {A.~D.}\ \bibnamefont {Bershadsky}}, \ and\ \bibinfo
  {author} {\bibfnamefont {M.~P.}\ \bibnamefont {Sheetz}},\ }\href@noop {}
  {\bibfield  {journal} {\bibinfo  {journal} {PLoS ONE}\ }\textbf {\bibinfo
  {volume} {11}},\ \bibinfo {pages} {1} (\bibinfo {year} {2016})}\BibitemShut
  {NoStop}%
\bibitem [{\citenamefont {Bosk}\ \emph {et~al.}(2011)\citenamefont {Bosk},
  \citenamefont {Braunger}, \citenamefont {Gerke},\ and\ \citenamefont
  {Steinem}}]{bosk_2011_ezrin}%
  \BibitemOpen
  \bibfield  {author} {\bibinfo {author} {\bibfnamefont {S.}~\bibnamefont
  {Bosk}}, \bibinfo {author} {\bibfnamefont {J.~A.}\ \bibnamefont {Braunger}},
  \bibinfo {author} {\bibfnamefont {V.}~\bibnamefont {Gerke}}, \ and\ \bibinfo
  {author} {\bibfnamefont {C.}~\bibnamefont {Steinem}},\ }\href@noop {}
  {\bibfield  {journal} {\bibinfo  {journal} {Biophys. J.}\ }\textbf {\bibinfo
  {volume} {100}},\ \bibinfo {pages} {1708} (\bibinfo {year}
  {2011})}\BibitemShut {NoStop}%
\bibitem [{\citenamefont {Barnhart}\ \emph {et~al.}(2011)\citenamefont
  {Barnhart}, \citenamefont {Lee}, \citenamefont {Keren}, \citenamefont
  {Mogilner},\ and\ \citenamefont
  {Theriot}}]{barnhart_adhesion-dependent_2011}%
  \BibitemOpen
  \bibfield  {author} {\bibinfo {author} {\bibfnamefont {E.}~\bibnamefont
  {Barnhart}}, \bibinfo {author} {\bibfnamefont {K.}~\bibnamefont {Lee}},
  \bibinfo {author} {\bibfnamefont {K.}~\bibnamefont {Keren}}, \bibinfo
  {author} {\bibfnamefont {A.}~\bibnamefont {Mogilner}}, \ and\ \bibinfo
  {author} {\bibfnamefont {J.}~\bibnamefont {Theriot}},\ }\href@noop {}
  {\bibfield  {journal} {\bibinfo  {journal} {PLoS Biol.}\ }\textbf {\bibinfo
  {volume} {9}},\ \bibinfo {pages} {e1001059} (\bibinfo {year}
  {2011})}\BibitemShut {NoStop}%
\bibitem [{\citenamefont {Jurado}\ \emph {et~al.}(2005)\citenamefont {Jurado},
  \citenamefont {Haserick},\ and\ \citenamefont
  {Lee}}]{jurado_retrograde_2005}%
  \BibitemOpen
  \bibfield  {author} {\bibinfo {author} {\bibfnamefont {C.}~\bibnamefont
  {Jurado}}, \bibinfo {author} {\bibfnamefont {J.~R.}\ \bibnamefont
  {Haserick}}, \ and\ \bibinfo {author} {\bibfnamefont {J.}~\bibnamefont
  {Lee}},\ }\href@noop {} {\bibfield  {journal} {\bibinfo  {journal} {Mol.
  Biol. Cell}\ }\textbf {\bibinfo {volume} {16}},\ \bibinfo {pages} {507}
  (\bibinfo {year} {2005})}\BibitemShut {NoStop}%
\bibitem [{Note1()}]{Note1}%
  \BibitemOpen
  \bibinfo {note} {It is important to note that the universal relation between
  shape and motility is only obtained when the flexocyte velocity multiplied by
  the substrate friction is employed as relevant observable, not for the
  flexocyte velocity. The relevant observable is therefore the instantaneous
  total propulsion force, $F_{act}=\protect \mathaccentV {tilde}07E{v}\protect
  \mathaccentV {tilde}07E{\gamma }_{\protect \textrm {s}}$.}\BibitemShut
  {Stop}%
\bibitem [{\citenamefont {Golestanian}\ \emph {et~al.}(2005)\citenamefont
  {Golestanian}, \citenamefont {Liverpool},\ and\ \citenamefont
  {Ajdari}}]{golestanian_propulsion_2005}%
  \BibitemOpen
  \bibfield  {author} {\bibinfo {author} {\bibfnamefont {R.}~\bibnamefont
  {Golestanian}}, \bibinfo {author} {\bibfnamefont {T.}~\bibnamefont
  {Liverpool}}, \ and\ \bibinfo {author} {\bibfnamefont {A.}~\bibnamefont
  {Ajdari}},\ }\href@noop {} {\bibfield  {journal} {\bibinfo  {journal} {Phys.
  Rev. Lett.}\ }\textbf {\bibinfo {volume} {94}},\ \bibinfo {pages} {220801}
  (\bibinfo {year} {2005})}\BibitemShut {NoStop}%
\bibitem [{Note2()}]{Note2}%
  \BibitemOpen
  \bibinfo {note} {Positive values of $\Delta \theta $ indicate a deflection
  towards the $x$-axis and negative values indicate a deflection towards the
  $y$-axis.}\BibitemShut {Stop}%
\bibitem [{\citenamefont {Tojkander}\ \emph {et~al.}(2012)\citenamefont
  {Tojkander}, \citenamefont {Gateva},\ and\ \citenamefont
  {Lappalainen}}]{tojkander_actin_2012}%
  \BibitemOpen
  \bibfield  {author} {\bibinfo {author} {\bibfnamefont {S.}~\bibnamefont
  {Tojkander}}, \bibinfo {author} {\bibfnamefont {G.}~\bibnamefont {Gateva}}, \
  and\ \bibinfo {author} {\bibfnamefont {P.}~\bibnamefont {Lappalainen}},\
  }\href@noop {} {\bibfield  {journal} {\bibinfo  {journal} {J. Cell Sci.}\
  }\textbf {\bibinfo {volume} {125}},\ \bibinfo {pages} {1855} (\bibinfo {year}
  {2012})}\BibitemShut {NoStop}%
\bibitem [{\citenamefont {Hotulainen}\ and\ \citenamefont
  {Lappalainen}(2006)}]{hotulainen_fiber_2006}%
  \BibitemOpen
  \bibfield  {author} {\bibinfo {author} {\bibfnamefont {P.}~\bibnamefont
  {Hotulainen}}\ and\ \bibinfo {author} {\bibfnamefont {P.}~\bibnamefont
  {Lappalainen}},\ }\href@noop {} {\bibfield  {journal} {\bibinfo  {journal}
  {J. Cell Biol.}\ }\textbf {\bibinfo {volume} {173}},\ \bibinfo {pages} {383}
  (\bibinfo {year} {2006})}\BibitemShut {NoStop}%
\bibitem [{\citenamefont {Barnhart}\ \emph {et~al.}(2015)\citenamefont
  {Barnhart}, \citenamefont {Lee}, \citenamefont {Allen}, \citenamefont
  {Theriot},\ and\ \citenamefont {Mogilner}}]{barnhart_balance_2015}%
  \BibitemOpen
  \bibfield  {author} {\bibinfo {author} {\bibfnamefont {E.}~\bibnamefont
  {Barnhart}}, \bibinfo {author} {\bibfnamefont {K.}~\bibnamefont {Lee}},
  \bibinfo {author} {\bibfnamefont {G.}~\bibnamefont {Allen}}, \bibinfo
  {author} {\bibfnamefont {J.}~\bibnamefont {Theriot}}, \ and\ \bibinfo
  {author} {\bibfnamefont {A.}~\bibnamefont {Mogilner}},\ }\href@noop {}
  {\bibfield  {journal} {\bibinfo  {journal} {Proc. Natl. Acad. Sci. U.S.A.}\
  }\textbf {\bibinfo {volume} {112}},\ \bibinfo {pages} {5045} (\bibinfo {year}
  {2015})}\BibitemShut {NoStop}%
\bibitem [{\citenamefont {Batchelder}\ \emph {et~al.}(2011)\citenamefont
  {Batchelder}, \citenamefont {Hollopeter}, \citenamefont {Campillo},
  \citenamefont {Mezanges}, \citenamefont {Jorgensen}, \citenamefont {Nassoy},
  \citenamefont {Sens},\ and\ \citenamefont
  {Plastino}}]{batchelder_tension_2011}%
  \BibitemOpen
  \bibfield  {author} {\bibinfo {author} {\bibfnamefont {E.~L.}\ \bibnamefont
  {Batchelder}}, \bibinfo {author} {\bibfnamefont {G.}~\bibnamefont
  {Hollopeter}}, \bibinfo {author} {\bibfnamefont {C.}~\bibnamefont
  {Campillo}}, \bibinfo {author} {\bibfnamefont {X.}~\bibnamefont {Mezanges}},
  \bibinfo {author} {\bibfnamefont {E.~M.}\ \bibnamefont {Jorgensen}}, \bibinfo
  {author} {\bibfnamefont {P.}~\bibnamefont {Nassoy}}, \bibinfo {author}
  {\bibfnamefont {P.}~\bibnamefont {Sens}}, \ and\ \bibinfo {author}
  {\bibfnamefont {J.}~\bibnamefont {Plastino}},\ }\href@noop {} {\bibfield
  {journal} {\bibinfo  {journal} {Proc. Natl. Acad. Sci. U.S.A.}\ }\textbf
  {\bibinfo {volume} {108}},\ \bibinfo {pages} {11429} (\bibinfo {year}
  {2011})}\BibitemShut {NoStop}%
\bibitem [{\citenamefont {Elgeti}\ and\ \citenamefont
  {Gompper}(2013)}]{elgeti_surface_2013}%
  \BibitemOpen
  \bibfield  {author} {\bibinfo {author} {\bibfnamefont {J.}~\bibnamefont
  {Elgeti}}\ and\ \bibinfo {author} {\bibfnamefont {G.}~\bibnamefont
  {Gompper}},\ }\href@noop {} {\bibfield  {journal} {\bibinfo  {journal} {EPL
  (Europhys. Lett.)}\ }\textbf {\bibinfo {volume} {101}},\ \bibinfo {pages}
  {48003} (\bibinfo {year} {2013})}\BibitemShut {NoStop}%
\bibitem [{\citenamefont {Takatori}\ \emph {et~al.}(2014)\citenamefont
  {Takatori}, \citenamefont {Yan},\ and\ \citenamefont
  {Brady}}]{brady_pressure_2014}%
  \BibitemOpen
  \bibfield  {author} {\bibinfo {author} {\bibfnamefont {S.~C.}\ \bibnamefont
  {Takatori}}, \bibinfo {author} {\bibfnamefont {W.}~\bibnamefont {Yan}}, \
  and\ \bibinfo {author} {\bibfnamefont {J.~F.}\ \bibnamefont {Brady}},\
  }\href@noop {} {\bibfield  {journal} {\bibinfo  {journal} {Phys. Rev. Lett.}\
  }\textbf {\bibinfo {volume} {113}},\ \bibinfo {pages} {028103} (\bibinfo
  {year} {2014})}\BibitemShut {NoStop}%
\bibitem [{\citenamefont {Miyoshi}\ \emph {et~al.}(2012)\citenamefont
  {Miyoshi}, \citenamefont {Adachi}, \citenamefont {Ju}, \citenamefont {Lee},
  \citenamefont {Cho}, \citenamefont {Ko}, \citenamefont {Uchida},\ and\
  \citenamefont {Yamagata}}]{miyoshi_characteristics_2012}%
  \BibitemOpen
  \bibfield  {author} {\bibinfo {author} {\bibfnamefont {H.}~\bibnamefont
  {Miyoshi}}, \bibinfo {author} {\bibfnamefont {T.}~\bibnamefont {Adachi}},
  \bibinfo {author} {\bibfnamefont {J.}~\bibnamefont {Ju}}, \bibinfo {author}
  {\bibfnamefont {S.}~\bibnamefont {Lee}}, \bibinfo {author} {\bibfnamefont
  {D.}~\bibnamefont {Cho}}, \bibinfo {author} {\bibfnamefont {J.}~\bibnamefont
  {Ko}}, \bibinfo {author} {\bibfnamefont {G.}~\bibnamefont {Uchida}}, \ and\
  \bibinfo {author} {\bibfnamefont {Y.}~\bibnamefont {Yamagata}},\ }\href@noop
  {} {\bibfield  {journal} {\bibinfo  {journal} {Biomater.}\ }\textbf {\bibinfo
  {volume} {33}},\ \bibinfo {pages} {395} (\bibinfo {year} {2012})}\BibitemShut
  {NoStop}%
\bibitem [{\citenamefont {Miyoshi}\ \emph {et~al.}(2010)\citenamefont
  {Miyoshi}, \citenamefont {Ju}, \citenamefont {Lee}, \citenamefont {Cho},
  \citenamefont {Ko}, \citenamefont {Yamagata},\ and\ \citenamefont
  {Adachi}}]{miyoshi_control_2010}%
  \BibitemOpen
  \bibfield  {author} {\bibinfo {author} {\bibfnamefont {H.}~\bibnamefont
  {Miyoshi}}, \bibinfo {author} {\bibfnamefont {J.}~\bibnamefont {Ju}},
  \bibinfo {author} {\bibfnamefont {S.}~\bibnamefont {Lee}}, \bibinfo {author}
  {\bibfnamefont {D.}~\bibnamefont {Cho}}, \bibinfo {author} {\bibfnamefont
  {J.}~\bibnamefont {Ko}}, \bibinfo {author} {\bibfnamefont {Y.}~\bibnamefont
  {Yamagata}}, \ and\ \bibinfo {author} {\bibfnamefont {T.}~\bibnamefont
  {Adachi}},\ }\href@noop {} {\bibfield  {journal} {\bibinfo  {journal}
  {Biomater.}\ }\textbf {\bibinfo {volume} {31}},\ \bibinfo {pages} {8539}
  (\bibinfo {year} {2010})}\BibitemShut {NoStop}%
\bibitem [{\citenamefont {Lieber}\ \emph {et~al.}(2013)\citenamefont {Lieber},
  \citenamefont {Yehudai-Resheff}, \citenamefont {Barnhart}, \citenamefont
  {Theriot},\ and\ \citenamefont {Keren}}]{lieber_tension_2013}%
  \BibitemOpen
  \bibfield  {author} {\bibinfo {author} {\bibfnamefont {A.~D.}\ \bibnamefont
  {Lieber}}, \bibinfo {author} {\bibfnamefont {S.}~\bibnamefont
  {Yehudai-Resheff}}, \bibinfo {author} {\bibfnamefont {E.~L.}\ \bibnamefont
  {Barnhart}}, \bibinfo {author} {\bibfnamefont {J.~A.}\ \bibnamefont
  {Theriot}}, \ and\ \bibinfo {author} {\bibfnamefont {K.}~\bibnamefont
  {Keren}},\ }\href@noop {} {\bibfield  {journal} {\bibinfo  {journal} {Curr.
  Biol.}\ }\textbf {\bibinfo {volume} {23}},\ \bibinfo {pages} {1409} (\bibinfo
  {year} {2013})}\BibitemShut {NoStop}%
\bibitem [{\citenamefont {Turlier}\ \emph {et~al.}(2016)\citenamefont
  {Turlier}, \citenamefont {Fedosov}, \citenamefont {Audoly}, \citenamefont
  {Auth}, \citenamefont {Gov}, \citenamefont {Sykes}, \citenamefont {Joanny},
  \citenamefont {Gompper},\ and\ \citenamefont
  {Betz}}]{betz_fluctuations_2016}%
  \BibitemOpen
  \bibfield  {author} {\bibinfo {author} {\bibfnamefont {H.}~\bibnamefont
  {Turlier}}, \bibinfo {author} {\bibfnamefont {D.~A.}\ \bibnamefont
  {Fedosov}}, \bibinfo {author} {\bibfnamefont {B.}~\bibnamefont {Audoly}},
  \bibinfo {author} {\bibfnamefont {T.}~\bibnamefont {Auth}}, \bibinfo {author}
  {\bibfnamefont {N.~S.}\ \bibnamefont {Gov}}, \bibinfo {author} {\bibfnamefont
  {C.}~\bibnamefont {Sykes}}, \bibinfo {author} {\bibfnamefont {J.-F.}\
  \bibnamefont {Joanny}}, \bibinfo {author} {\bibfnamefont {G.}~\bibnamefont
  {Gompper}}, \ and\ \bibinfo {author} {\bibfnamefont {T.}~\bibnamefont
  {Betz}},\ }\href@noop {} {\bibfield  {journal} {\bibinfo  {journal} {Nat.
  Phys.}\ }\textbf {\bibinfo {volume} {12}},\ \bibinfo {pages} {513} (\bibinfo
  {year} {2016})}\BibitemShut {NoStop}%
\bibitem [{\citenamefont {Monzel}\ and\ \citenamefont
  {Sengupta}(2016)}]{monzel_fluctuations_2016}%
  \BibitemOpen
  \bibfield  {author} {\bibinfo {author} {\bibfnamefont {C.}~\bibnamefont
  {Monzel}}\ and\ \bibinfo {author} {\bibfnamefont {K.}~\bibnamefont
  {Sengupta}},\ }\href@noop {} {\bibfield  {journal} {\bibinfo  {journal} {J.
  Phys. D}\ }\textbf {\bibinfo {volume} {49}},\ \bibinfo {pages} {243002}
  (\bibinfo {year} {2016})}\BibitemShut {NoStop}%
\bibitem [{\citenamefont {Paoluzzi}\ \emph {et~al.}(2016)\citenamefont
  {Paoluzzi}, \citenamefont {DiLeonardo}, \citenamefont {Marchetti},\ and\
  \citenamefont {Angelani}}]{dileonardo_vesicle_2016}%
  \BibitemOpen
  \bibfield  {author} {\bibinfo {author} {\bibfnamefont {M.}~\bibnamefont
  {Paoluzzi}}, \bibinfo {author} {\bibfnamefont {R.}~\bibnamefont
  {DiLeonardo}}, \bibinfo {author} {\bibfnamefont {M.}~\bibnamefont
  {Marchetti}}, \ and\ \bibinfo {author} {\bibfnamefont {L.}~\bibnamefont
  {Angelani}},\ }\href@noop {} {\bibfield  {journal} {\bibinfo  {journal} {Sci.
  Rep.}\ }\textbf {\bibinfo {volume} {6}},\ \bibinfo {pages} {34146} (\bibinfo
  {year} {2016})}\BibitemShut {NoStop}%
\bibitem [{\citenamefont {Loubet}\ \emph {et~al.}(2012)\citenamefont {Loubet},
  \citenamefont {Seifert},\ and\ \citenamefont {Lomholt}}]{loubet1203}%
  \BibitemOpen
  \bibfield  {author} {\bibinfo {author} {\bibfnamefont {B.}~\bibnamefont
  {Loubet}}, \bibinfo {author} {\bibfnamefont {U.}~\bibnamefont {Seifert}}, \
  and\ \bibinfo {author} {\bibfnamefont {M.~A.}\ \bibnamefont {Lomholt}},\
  }\href@noop {} {\bibfield  {journal} {\bibinfo  {journal} {Phys. Rev. E}\
  }\textbf {\bibinfo {volume} {85}},\ \bibinfo {pages} {031913} (\bibinfo
  {year} {2012})}\BibitemShut {NoStop}%
\bibitem [{\citenamefont {Br{\"u}ckner}\ \emph {et~al.}(2019)\citenamefont
  {Br{\"u}ckner}, \citenamefont {Fink}, \citenamefont {Schreiber},
  \citenamefont {R{\"o}ttgermann}, \citenamefont {R{\"a}dler},\ and\
  \citenamefont {Broedersz}}]{bruckner1903}%
  \BibitemOpen
  \bibfield  {author} {\bibinfo {author} {\bibfnamefont {D.~B.}\ \bibnamefont
  {Br{\"u}ckner}}, \bibinfo {author} {\bibfnamefont {A.}~\bibnamefont {Fink}},
  \bibinfo {author} {\bibfnamefont {C.}~\bibnamefont {Schreiber}}, \bibinfo
  {author} {\bibfnamefont {P.~J.}\ \bibnamefont {R{\"o}ttgermann}}, \bibinfo
  {author} {\bibfnamefont {J.~O.}\ \bibnamefont {R{\"a}dler}}, \ and\ \bibinfo
  {author} {\bibfnamefont {C.~P.}\ \bibnamefont {Broedersz}},\ }\href@noop {}
  {\bibfield  {journal} {\bibinfo  {journal} {Nat. Phys.}\ }\textbf {\bibinfo
  {volume} {15}},\ \bibinfo {pages} {595} (\bibinfo {year} {2019})}\BibitemShut
  {NoStop}%
\bibitem [{\citenamefont {Nanba}\ \emph {et~al.}(2015)\citenamefont {Nanba},
  \citenamefont {Toki}, \citenamefont {Tate}, \citenamefont {Imai},
  \citenamefont {Matsushita}, \citenamefont {Shiraishi}, \citenamefont
  {Sayama}, \citenamefont {Toki}, \citenamefont {Higashiyama},\ and\
  \citenamefont {Barrandon}}]{nanba_rotation_2015}%
  \BibitemOpen
  \bibfield  {author} {\bibinfo {author} {\bibfnamefont {D.}~\bibnamefont
  {Nanba}}, \bibinfo {author} {\bibfnamefont {F.}~\bibnamefont {Toki}},
  \bibinfo {author} {\bibfnamefont {S.}~\bibnamefont {Tate}}, \bibinfo {author}
  {\bibfnamefont {M.}~\bibnamefont {Imai}}, \bibinfo {author} {\bibfnamefont
  {N.}~\bibnamefont {Matsushita}}, \bibinfo {author} {\bibfnamefont
  {K.}~\bibnamefont {Shiraishi}}, \bibinfo {author} {\bibfnamefont
  {K.}~\bibnamefont {Sayama}}, \bibinfo {author} {\bibfnamefont
  {H.}~\bibnamefont {Toki}}, \bibinfo {author} {\bibfnamefont {S.}~\bibnamefont
  {Higashiyama}}, \ and\ \bibinfo {author} {\bibfnamefont {Y.}~\bibnamefont
  {Barrandon}},\ }\href@noop {} {\bibfield  {journal} {\bibinfo  {journal} {J.
  Cell Biol.}\ }\textbf {\bibinfo {volume} {209}},\ \bibinfo {pages} {305}
  (\bibinfo {year} {2015})}\BibitemShut {NoStop}%
\bibitem [{\citenamefont {Banerjee}\ \emph {et~al.}(2015)\citenamefont
  {Banerjee}, \citenamefont {Utuje},\ and\ \citenamefont
  {Marchetti}}]{banerjee1506}%
  \BibitemOpen
  \bibfield  {author} {\bibinfo {author} {\bibfnamefont {S.}~\bibnamefont
  {Banerjee}}, \bibinfo {author} {\bibfnamefont {K.~J.}\ \bibnamefont {Utuje}},
  \ and\ \bibinfo {author} {\bibfnamefont {M.~C.}\ \bibnamefont {Marchetti}},\
  }\href@noop {} {\bibfield  {journal} {\bibinfo  {journal} {Phys. Rev. Lett.}\
  }\textbf {\bibinfo {volume} {114}},\ \bibinfo {pages} {228101} (\bibinfo
  {year} {2015})}\BibitemShut {NoStop}%
\bibitem [{\citenamefont {Gross}\ \emph {et~al.}(2019)\citenamefont {Gross},
  \citenamefont {Kumar}, \citenamefont {Goehring}, \citenamefont {Bois},
  \citenamefont {Hoege}, \citenamefont {J{\"u}licher},\ and\ \citenamefont
  {Grill}}]{gross1903}%
  \BibitemOpen
  \bibfield  {author} {\bibinfo {author} {\bibfnamefont {P.}~\bibnamefont
  {Gross}}, \bibinfo {author} {\bibfnamefont {K.~V.}\ \bibnamefont {Kumar}},
  \bibinfo {author} {\bibfnamefont {N.~W.}\ \bibnamefont {Goehring}}, \bibinfo
  {author} {\bibfnamefont {J.~S.}\ \bibnamefont {Bois}}, \bibinfo {author}
  {\bibfnamefont {C.}~\bibnamefont {Hoege}}, \bibinfo {author} {\bibfnamefont
  {F.}~\bibnamefont {J{\"u}licher}}, \ and\ \bibinfo {author} {\bibfnamefont
  {S.~W.}\ \bibnamefont {Grill}},\ }\href@noop {} {\bibfield  {journal}
  {\bibinfo  {journal} {Nat. Phys.}\ }\textbf {\bibinfo {volume} {15}},\
  \bibinfo {pages} {293} (\bibinfo {year} {2019})}\BibitemShut {NoStop}%
\bibitem [{\citenamefont {Fritz-Laylin}\ \emph {et~al.}(2017)\citenamefont
  {Fritz-Laylin}, \citenamefont {Riel-Mehan}, \citenamefont {Chen},
  \citenamefont {Lord}, \citenamefont {Goddard}, \citenamefont {Ferrin},
  \citenamefont {Nicholson-Dykstra}, \citenamefont {Higgs}, \citenamefont
  {Johnson}, \citenamefont {Betzig} \emph {et~al.}}]{fritz-laylin17}%
  \BibitemOpen
  \bibfield  {author} {\bibinfo {author} {\bibfnamefont {L.~K.}\ \bibnamefont
  {Fritz-Laylin}}, \bibinfo {author} {\bibfnamefont {M.}~\bibnamefont
  {Riel-Mehan}}, \bibinfo {author} {\bibfnamefont {B.-C.}\ \bibnamefont
  {Chen}}, \bibinfo {author} {\bibfnamefont {S.~J.}\ \bibnamefont {Lord}},
  \bibinfo {author} {\bibfnamefont {T.~D.}\ \bibnamefont {Goddard}}, \bibinfo
  {author} {\bibfnamefont {T.~E.}\ \bibnamefont {Ferrin}}, \bibinfo {author}
  {\bibfnamefont {S.~M.}\ \bibnamefont {Nicholson-Dykstra}}, \bibinfo {author}
  {\bibfnamefont {H.}~\bibnamefont {Higgs}}, \bibinfo {author} {\bibfnamefont
  {G.~T.}\ \bibnamefont {Johnson}}, \bibinfo {author} {\bibfnamefont
  {E.}~\bibnamefont {Betzig}},  \emph {et~al.},\ }\href@noop {} {\bibfield
  {journal} {\bibinfo  {journal} {eLife}\ }\textbf {\bibinfo {volume} {6}},\
  \bibinfo {pages} {e26990} (\bibinfo {year} {2017})}\BibitemShut {NoStop}%
\bibitem [{\citenamefont {Whitesides}(2018)}]{whitesides1803}%
  \BibitemOpen
  \bibfield  {author} {\bibinfo {author} {\bibfnamefont {G.~M.}\ \bibnamefont
  {Whitesides}},\ }\href@noop {} {\bibfield  {journal} {\bibinfo  {journal}
  {Angew. Chem. Int. Ed.}\ }\textbf {\bibinfo {volume} {57}},\ \bibinfo {pages}
  {4258} (\bibinfo {year} {2018})}\BibitemShut {NoStop}%
\bibitem [{\citenamefont {Horowitz}\ \emph {et~al.}(2019)\citenamefont
  {Horowitz}, \citenamefont {Chambers}, \citenamefont {G{\"o}zen},
  \citenamefont {Dimiduk},\ and\ \citenamefont {Manoharan}}]{horowitz1806}%
  \BibitemOpen
  \bibfield  {author} {\bibinfo {author} {\bibfnamefont {V.~R.}\ \bibnamefont
  {Horowitz}}, \bibinfo {author} {\bibfnamefont {Z.~C.}\ \bibnamefont
  {Chambers}}, \bibinfo {author} {\bibfnamefont {I.}~\bibnamefont {G{\"o}zen}},
  \bibinfo {author} {\bibfnamefont {T.~G.}\ \bibnamefont {Dimiduk}}, \ and\
  \bibinfo {author} {\bibfnamefont {V.~N.}\ \bibnamefont {Manoharan}},\
  }\href@noop {} {\bibfield  {journal} {\bibinfo  {journal} {Eur. Phys. J.:
  Spec. Top.}\ }\textbf {\bibinfo {volume} {227}},\ \bibinfo {pages} {2413}
  (\bibinfo {year} {2019})}\BibitemShut {NoStop}%
\end{thebibliography}%


\begin{thebibliography}{0}%
\makeatletter
\providecommand \@ifxundefined [1]{%
 \@ifx{#1\undefined}
}%
\providecommand \@ifnum [1]{%
 \ifnum #1\expandafter \@firstoftwo
 \else \expandafter \@secondoftwo
 \fi
}%
\providecommand \@ifx [1]{%
 \ifx #1\expandafter \@firstoftwo
 \else \expandafter \@secondoftwo
 \fi
}%
\providecommand \natexlab [1]{#1}%
\providecommand \enquote  [1]{``#1''}%
\providecommand \bibnamefont  [1]{#1}%
\providecommand \bibfnamefont [1]{#1}%
\providecommand \citenamefont [1]{#1}%
\providecommand \href@noop [0]{\@secondoftwo}%
\providecommand \href [0]{\begingroup \@sanitize@url \@href}%
\providecommand \@href[1]{\@@startlink{#1}\@@href}%
\providecommand \@@href[1]{\endgroup#1\@@endlink}%
\providecommand \@sanitize@url [0]{\catcode `\\12\catcode `\$12\catcode
  `\&12\catcode `\#12\catcode `\^12\catcode `\_12\catcode `\%12\relax}%
\providecommand \@@startlink[1]{}%
\providecommand \@@endlink[0]{}%
\providecommand \url  [0]{\begingroup\@sanitize@url \@url }%
\providecommand \@url [1]{\endgroup\@href {#1}{\urlprefix }}%
\providecommand \urlprefix  [0]{URL }%
\providecommand \Eprint [0]{\href }%
\providecommand \doibase [0]{http://dx.doi.org/}%
\providecommand \selectlanguage [0]{\@gobble}%
\providecommand \bibinfo  [0]{\@secondoftwo}%
\providecommand \bibfield  [0]{\@secondoftwo}%
\providecommand \translation [1]{[#1]}%
\providecommand \BibitemOpen [0]{}%
\providecommand \bibitemStop [0]{}%
\providecommand \bibitemNoStop [0]{.\EOS\space}%
\providecommand \EOS [0]{\spacefactor3000\relax}%
\providecommand \BibitemShut  [1]{\csname bibitem#1\endcsname}%
\let\auto@bib@innerbib\@empty
\end{thebibliography}%
    
\end{document}


\title{Self-organized motility of vesicles with internal active filaments \\ Supporting Information}

 \author{Clara Abaurrea-Velasco}
    \affiliation{
        Theoretical Soft Matter and Biophysics, Institute of Complex Systems and Institute for Advanced Simulation, Forschungszentrum J\"ulich, D-52425 J\"ulich, Germany
    }
    \email{c.abaurreavelasco@uu.nl; t.auth@fz-juelich.de; g.gompper@fz-juelich.de}

   \author{Thorsten Auth}
    \affiliation{
    Theoretical Soft Matter and Biophysics, Institute of Complex Systems and Institute for Advanced Simulation, Forschungszentrum J\"ulich, D-52425 J\"ulich, Germany
    }

    \author{Gerhard Gompper}
    \affiliation{
    Theoretical Soft Matter and Biophysics, Institute of Complex Systems and Institute for Advanced Simulation, Forschungszentrum J\"ulich, D-52425 J\"ulich, Germany
    }

 
\maketitle

\section{DEFLECTION AND SHAPE RELAXATION AT FRICTION INTERFACES}

\begin{figure*}
        \centering
        \includegraphics[width=1.9\columnwidth]{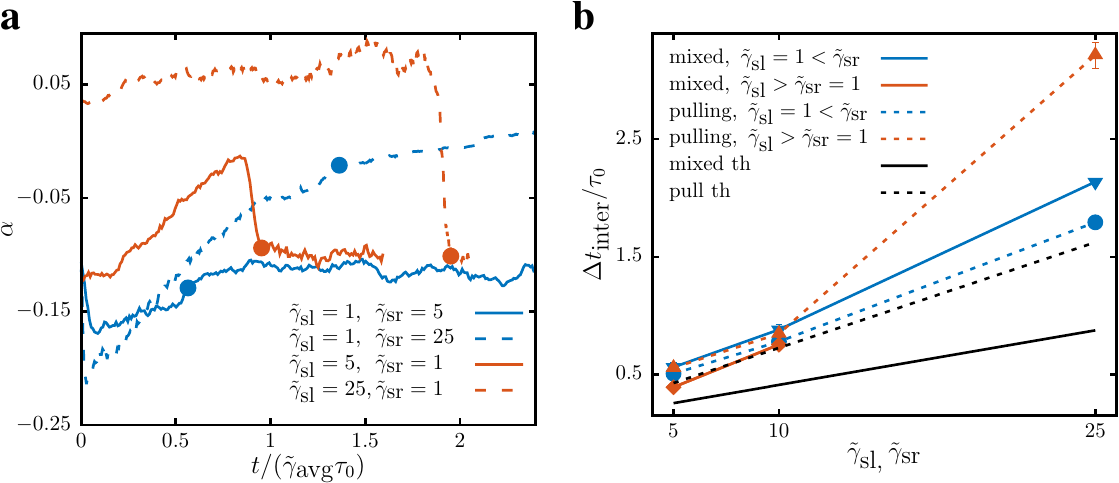}
        \caption{Flexocytes at friction interfaces. Dynamics and relaxation of shape changes for various substrate frictions $\tilde{\gamma}_\textrm{sl},\,\tilde{\gamma}_\textrm{sr}$. \textbf{a)} Time dependence of the signed asphericity $\alpha$, where $\tilde{\gamma}_\textrm{avg}=(\tilde{\gamma}_\textrm{sl}+\tilde{\gamma}_\textrm{sr})/2$. Flexocytes of puller type with $N_\textrm{r,push}=N_\textrm{r,pull}=64$, $\textrm{Pe}=100$, $\tilde{k}_A=100$ for $\theta_\textrm{i}=90\textordmasculine$ and various $\tilde{\gamma}_{\textrm{s,l}}$ and $\tilde{\gamma}_{\textrm{s,r}}$. $t=0$ coincides with the time when the polymer ring first comes in contact with the interface. The points indicate the time when the dorsal end crosses the interface.
       \textbf{b)} Time needed for flexocytes to cross the friction interfaces $\Delta t_\textrm{inter}$ versus substrate frictions $\tilde{\gamma}_{\textrm{sl}}$ and $\tilde{\gamma}_{\textrm{sr}}$ for normal angle of incidence $\theta_{\rm i}\approx90\textordmasculine$. Flexocytes of mixed type with $N_\textrm{r,push}=N_\textrm{r,pull}=64$, $\textrm{Pe}=100$, $\tilde{k}_A=100$, and corresponding data for flexocytes of puller type. Systems where $\tilde{\gamma}_{\textrm{sl}}=1$, $\tilde{\gamma}_{\textrm{sr}}$ varies and vice-versa. No data exists for the flexocytes of mixed type for $\tilde{\gamma}_\textrm{sl}>\tilde{\gamma}_\textrm{sr}$ and $\tilde{\gamma}_\textrm{s}=25$. Theoretical curves are indicated by black dashed lines. See movies M12 to M15.
}
        \label{fig:ang_inter2}
\end{figure*}


The dynamics of transient and persistent shape changes of flexocytes at friction interfaces reflects the interplay of shape deformation and internal reorganization of the filaments, see Fig.~\ref{fig:ang_inter2}(a).
$K$-flexocytes of puller type flatten ($\Delta \alpha <0$) and elongate along their long axis for a transition from a small-friction to a large-friction, $\tilde{\gamma}_{\textrm{sl}}<\tilde{\gamma}_{\textrm{sl}}$, substrate because of the higher velocity at the dorsal end compared with the apical end. Vice versa, they round up for a transition from a large-friction to a small-friction substrate because of the higher velocity at the apical end compared with the dorsal end.

For a permanent change of the stationary flexocyte shape between "$N$" and "$K$" the dynamics strongly differs for opposite friction changes, see Fig.~\ref{fig:ang_inter2}(a). The shape of an $N$-flexocyte remains almost unchanged until the flexocyte has completely transitioned to the region with smaller substrate friction and decreases rapidly once the pulling filaments reach the new friction region. 
In contrast, a $K$-flexocyte that moves across an interface from small to large friction first flattens (large negative asphericity) as soon as it touches the interface, then elongates (peak of positive sphericity) to become an $N$-flexocyte as it transitions into the large-friction region, see Fig.~\ref{fig:ang_inter2}(a).
Because of the larger final friction, the relaxation to the new final shape also lasts much longer than for the opposite process.

%
%
Figure~\ref{fig:ang_inter2}(b) shows the time $\Delta t_\textrm{inter}$ needed for a flexocyte to cross a friction interface for
various conditions.
These times can be compared with a simple estimate
\begin{equation}
\Delta t_\textrm{inter}=2R_\textrm{m}(N_\textrm{r}\gamma_{\text{r}\parallel} + n_\textrm{b} (\gamma_\textrm{sl}+\gamma_\textrm{sr})/2) / (N_\textrm{r}|\mathbf{F}_{\text{p}}|)
\end{equation}
obtained for a circular shape with all filaments oriented normal to the interface.
The interface-crossing times obtained in the simulations are typically longer than the simple estimate. This
results from smaller propulsion velocities because not all filaments are aligned in the direction of motion of the flexocyte,
and from parts of trajectories parallel to the interface.
In particular for $N$-flexocytes of mixed type, $\Delta t_\textrm{inter}$ is significantly longer
than the estimate because only very few pulling filaments point in the direction of motion, see Fig.~2 in the main text.
The sign of friction change also strongly affects the time scale of shape relaxation.
For motion from small to large friction, the $K$-flexocyte first flattens as the apical membrane reaches the interface and then
slowly relaxes to a more rounded shape of an $N$-flexocyte (for $\tilde{\gamma}_{\textrm{sr}}=25$) as the apical part moves into the
large-friction region.  In contrast, for motion from large to small substrate friction, the $N$-flexocyte shape remains almost unchanged
until the flexocyte has completely transitioned into the small-friction region and then relaxes rapidly to a $K$-flexocyte.

\section{SUPPLEMETARY DATA}
\CLA{Tables \ref{tab:mix} and \ref{tab:pull} contain parameter values and resulting shapes for the flexocytes presented in the scatter plots in Figs.~2 (c) and (d).}

\section{SUPPLEMETARY MOVIES}
\textbf{M1}: $F$-flexocyte of mixed type with $N_\textrm{r, push}=N_\textrm{r, pull}=16$, $\textrm{Pe}=25$, $\tilde{k}_A=100$, $\tilde{\gamma}_\textrm{s}=5$, and time elapsed between frames $dt/\tau_0=0.5$. The left side of the video shows the dynamics of the semiflexible ring with attached self-propelled filaments, the right side shows the trajectory of the ring center-of-mass, see Fig.~2 in the main text.

\textbf{M2-A}: $K$-flexocyte of mixed type with $N_\textrm{r, push}=N_\textrm{r, pull}=40$, $\textrm{Pe}=50$, $\tilde{k}_A=1$, $\tilde{\gamma}_\textrm{s}=1$, and time elapsed between frames $dt/\tau_0=0.1$. The left side of the video shows the dynamics of the semiflexible ring with attached self-propelled filaments, the right side shows the trajectory of the ring center-of-mass, see Fig.~2 in the main text.

\textbf{M2-B}: $K$-flexocyte of mixed type with $N_\textrm{r, push}=N_\textrm{r, pull}=64$, $\textrm{Pe}=25$, $\tilde{k}_A=1$, $\tilde{\gamma}_\textrm{s}=1$, and time elapsed between frames $dt/\tau_0=0.5$. The flexocytes performs circling motion. The left side of the video shows the dynamics of the semiflexible ring with attached self-propelled filaments, the right side shows the trajectory of the ring center-of-mass, see Fig.~2 in the main text.

\textbf{M3}: $N_{mono}$-flexocyte of mixed type with $N_\textrm{r, push}=N_\textrm{r, pull}=40$, $\textrm{Pe}=50$, $\tilde{k}_A=1$, $\tilde{\gamma}_\textrm{s}=10$, and time elapsed between frames $dt/\tau_0=0.5$. The left side of the video shows the dynamics of the semiflexible ring with attached self-propelled filaments, the right side shows the trajectory of the ring center-of-mass, see Fig.~2 in the main text.

\textbf{M4}: $N_{bi}$-flexocyte of mixed type with $N_\textrm{r, push}=N_\textrm{r, pull}=64$, $\textrm{Pe}=100$, $\tilde{k}_A=100$, $\tilde{\gamma}_\textrm{s}=1$, and time elapsed between frames $dt/\tau_0=0.05$. The left side of the video shows the dynamics of the semiflexible ring with attached self-propelled filaments, the right side shows the trajectory of the ring center-of-mass, see Fig.~2 in the main text.

\textbf{M5}: $F$-flexocyte of puller type with $N_\textrm{r, pull}=16$, $\textrm{Pe}=25$, $\tilde{k}_A=100$, $\tilde{\gamma}_\textrm{s}=5$, and time elapsed between frames $dt/\tau_0=0.25$. The left side of the video shows the dynamics of the semiflexible ring with attached self-propelled filaments, the right side shows the trajectory of the ring center-of-mass, see Fig.~2 in the main text.

\textbf{M6}: $K$-flexocyte of puller type with $N_\textrm{r, pull}=40$, $\textrm{Pe}=50$, $\tilde{k}_A=1$, $\tilde{\gamma}_\textrm{s}=1$, and time elapsed between frames $dt/\tau_0=0.1$. The left side of the video shows the dynamics of the semiflexible ring with attached self-propelled filaments, the right side shows the trajectory of the ring center-of-mass, see Fig.~2 in the main text.

\textbf{M7}: $N$-flexocyte of puller type with $N_\textrm{r, pull}=64$, $\textrm{Pe}=50$, $\tilde{k}_A=1$, $\tilde{\gamma}_\textrm{s}=5$, and time elapsed between frames $dt/\tau_0=0.25$. The left side of the video shows the dynamics of the semiflexible ring with attached self-propelled filaments, the right side shows the trajectory of the ring center-of-mass, see Fig.~2 in the main text.

\textbf{M8}: Flexocyte interacting with a wall. $K$-flexocyte of mixed type with $N_\textrm{r, push}=N_\textrm{r, pull}=64$, $\textrm{Pe}=100$, $\tilde{k}_A=100$, $\tilde{\gamma}_\textrm{s}=1$, and time elapsed between frames $dt/\tau_0=0.1$. The wall is placed at $x=8R_\textrm{m}$, the angle of incidence is $\theta_{\rm i}\approx90\textordmasculine$ with respect to the y-axis. The left side of the video shows the dynamics of the semiflexible ring with attached self-propelled filaments, the right side shows the trajectory of the ring center-of-mass. The flexocyte meets the wall it deforms and it stays at the wall, see Fig.~4 in the main text. 

\textbf{M9}: Flexocyte interacting with a wall. $K$-flexocytes of mixed type with $N_\textrm{r, push}=N_\textrm{r, pull}=64$, $\textrm{Pe}=100$, $\tilde{k}_A=100$, $\tilde{\gamma}_\textrm{s}=1$, and time elapsed between frames $dt/\tau_0=0.1$. The wall is placed at $x=8R_\textrm{m}$, the angle of incidence is $\theta_{\rm i}\approx20\textordmasculine$ with respect to the y-axis. The left side of the video shows the dynamics of the semiflexible ring with attached self-propelled filaments, the right side shows the trajectory of the ring center-of-mass. The flexocyte meets the wall it deforms and it stays at the wall, see Fig.~4 in the main text.

\textbf{M10}: Flexocyte interacting with a wall. $K$-flexocyte of puller type with $N_\textrm{r, pull}=64$, $\textrm{Pe}=50$, $\tilde{k}_A=1$, $\tilde{\gamma}_\textrm{s}=1$, and time elapsed between frames $dt/\tau_0=0.1$. The wall is placed at $x=8R_\textrm{m}$, the angle of incidence is $\theta_{\rm i}\approx90\textordmasculine$ with respect to the y-axis. The left side of the video shows the dynamics of the semiflexible ring with attached self-propelled filaments, the right side shows the trajectory of the ring center-of-mass. The flexocyte meets the wall it deforms the filaments reorient and the flexocyte is reflected at a referred angle, see Fig.~4 in the main text.

\textbf{M11}: Flexocyte interacting with a wall. $K$-flexocyte of puller type with $N_\textrm{r, pull}=64$, $\textrm{Pe}=50$, $\tilde{k}_A=1$, $\tilde{\gamma}_\textrm{s}=1$, and time elapsed between frames $dt/\tau_0=0.1$. The wall is placed at $x=8R_\textrm{m}$, the angle of incidence is $\theta_{\rm i}\textrm{i}\approx20\textordmasculine$ with respect to the y-axis. The left side of the video shows the dynamics of the semiflexible ring with attached self-propelled filaments, the right side shows the trajectory of the ring center-of-mass. The flexocyte meets the wall it deforms the filaments reorient and the flexocyte is reflected at a referred angle, see Fig.~4 in the main text.

\textbf{M12}: Flexocyte at a friction interface. $N_{bi}$-flexocytes of mixed type with $N_\textrm{r, push}=N_\textrm{r, pull}=64$, $\textrm{Pe}=100$, $\tilde{k}_A=100$ and time elapsed between frames $dt/\tau_0=0.1$. The interface is placed at $x_\textrm{int}=3.5R_\textrm{m}$, the white region, $x<x_\textrm{int}$, has a substrate friction $\tilde{\gamma}_\textrm{sl}=1$, and the gray region, $x>x_\textrm{int}$, has a substrate friction $\tilde{\gamma}_\textrm{sr}=25$. The flexocyte is initialized as a neutrophil with two cluster of pushing filaments facing the interface with an angle of incidence $\theta_\textrm{i}\approx90\textordmasculine$. The left side of the video shows the dynamics of the semiflexible ring with attached self-propelled filaments, the right side shows the trajectory of the ring center-of-mass. The change in friction causes small transient changes in shape which lead to a reorientation of the filaments and a small deflection of the trajectory, see Fig.~5 in the main text.

\textbf{M13}: Flexocyte at a friction interface. $N_{bi}$-flexocytes of mixed type with $N_\textrm{r, push}=N_\textrm{r, pull}=64$, $\textrm{Pe}=100$, $\tilde{k}_A=100$ and time elapsed between frames $dt/\tau_0=0.1$. The interface is placed at $x_\textrm{int}=3.5R_\textrm{m}$, the gray region, $x<x_\textrm{int}$, has a substrate friction $\tilde{\gamma}_\textrm{sl}=5$, and the white region, $x>x_\textrm{int}$, has a substrate friction $\tilde{\gamma}_\textrm{sr}=1$. The flexocyte is initialized as a neutrophil with two cluster of pushing filaments facing the interface with an angle of incidence $\theta_\textrm{i}\approx30\textordmasculine$. The left side of the video shows the dynamics of the semiflexible ring with attached self-propelled filaments, the right side shows the trajectory of the ring center-of-mass. The change in friction causes small transient changes in shape which lead to a reorientation of the filaments and a small deflection of the trajectory, see Fig.~5 in the main text.

\textbf{M14}: Flexocyte at a friction interface. $K$-flexocytes of puller type wi $N_\textrm{r, pull}=64$, $\textrm{Pe}=100$, $\tilde{k}_A=100$ and time elapsed between frames $dt/\tau_0=0.1$. The interface is placed at $x_\textrm{int}=3.5R_\textrm{m}$, the white region, $x<x_\textrm{int}$, has a substrate friction $\tilde{\gamma}_\textrm{sl}=1$, and the gray region, $x>x_\textrm{int}$, has a substrate friction $\tilde{\gamma}_\textrm{sr}=25$. The flexocyte is initialized as a neutrophil with two cluster of pushing filaments facing the interface with an angle of incidence $\theta_\textrm{i}\approx90\textordmasculine$. The left side of the video shows the dynamics of the semiflexible ring with attached self-propelled filaments, the right side shows the trajectory of the ring center-of-mass. The change in friction causes transient and permanent shape changes, from a $K$-flexocytes to a $N$-flexocyte, which leads to a big deflection of the trajectory, see Fig.~5 in the main text.

\textbf{M15}: Flexocyte at a friction interface. $K$-flexocytes of puller type with $N_\textrm{r, pull}=64$, $\textrm{Pe}=100$, $\tilde{k}_A=100$ and time elapsed between frames $dt/\tau_0=0.1$. The interface is placed at $x_\textrm{int}=3.5R_\textrm{m}$, the  gray region, $x<x_\textrm{int}$, has a substrate friction $\tilde{\gamma}_\textrm{sl}=5$, and the white region, $x>x_\textrm{int}$, has a substrate friction $\tilde{\gamma}_\textrm{sr}=1$. The flexocyte is initialized as a neutrophil with two cluster of pushing filaments facing the interface with an angle of incidence $\theta_\textrm{i}\approx30\textordmasculine$. The left side of the video shows the dynamics of the semiflexible ring with attached self-propelled filaments, the right side shows the trajectory of the ring center-of-mass. The change in friction leads to a counter-clockwise tank treading of the semiflexible ring, which in turn orients the flexocyte parallel to the interface, see Fig.~5 in the main text.

\begin{table*}[]
	\centering
	\caption{Parameters for flexocytes of mixed type shown in Fig.~2 (c). For all the systems $\tilde{\kappa}=2$ and $\tilde{k}_{\rm S}=1000$.}\label{tab:mix}
	\rowcolors{2}{gray!25}{white}
	{\footnotesize
		\begin{tabular}{p{1cm}p{1cm}p{1cm}p{1cm}p{1cm}p{2cm}||p{1cm}p{1cm}p{1cm}p{1cm}p{1cm}p{2cm}}
			\hline
			$\tilde{k}_{\rm A}$ & $\tilde{\gamma}_{\rm s}$ & $N_\textrm{r,\textrm{push}}$ & $N_\textrm{r,\textrm{pull}}$ & Pe & Flexocyte type & $\tilde{k}_{\rm A}$ & $\tilde{\gamma}_{\rm s}$ & $N_\textrm{r,\textrm{push}}$ & $N_\textrm{r,\textrm{pull}}$ & Pe & Flexocyte type \\
			\hline
			\hline
			1 & 1 & 16 & 16 & 10  & $K$      & 100 & 1 & 16 & 16 & 10  & $K$ \\
			1 & 1 & 16 & 16 & 25  & $N_{bi}$ & 100 & 1 & 16 & 16 & 25  & $K$ \\
			1 & 1 & 16 & 16 & 50  & $K$      & 100 & 1 & 16 & 16 & 50  & $K$ \\
			1 & 1 & 16 & 16 & 100 & $K$      & 100 & 1 & 16 & 16 & 100 & $K$ \\
			1 & 1 & 40 & 40 & 10  & $K$      & 100 & 1 & 40 & 40 & 10  & $K$ \\
			1 & 1 & 40 & 40 & 25  & $N_{bi}$ & 100 & 1 & 40 & 40 & 25  & $K$ \\
			1 & 1 & 40 & 40 & 50  & $K$      & 100 & 1 & 40 & 40 & 50  & $K$ \\
			1 & 1 & 40 & 40 & 100 & $N_{bi}$ & 100 & 1 & 40 & 40 & 100 & $K$ \\
			1 & 1 & 64 & 64 & 10  & $K$      & 100 & 1 & 64 & 64 & 10  & $K$ \\
			1 & 1 & 64 & 64 & 25  & $N_{bi}$ & 100 & 1 & 64 & 64 & 25  & $N_{bi}$ \\
			1 & 1 & 64 & 64 & 50  & $N_{bi}$ & 100 & 1 & 64 & 64 & 50  & $F$ \\
			1 & 1 & 64 & 64 & 100 & $N_{bi}$ & 100 & 1 & 64 & 64 & 100 & $N_{bi}$ \\
			\hline	
			\hline	
			1 & 5 & 16 & 16 & 10  & $F$         & 100 & 5 & 16 & 16 & 10  & $F/N_{bi}$ \\
			1 & 5 & 16 & 16 & 25  & $N_{mono}$  & 100 & 5 & 16 & 16 & 25  & $F$ \\
			1 & 5 & 16 & 16 & 50  & $N_{mono}$  & 100 & 5 & 16 & 16 & 50  & $K$ \\
			1 & 5 & 16 & 16 & 100 & $K$         & 100 & 5 & 16 & 16 & 100 & $K$ \\
			1 & 5 & 40 & 40 & 10  & $N_{mono}$  & 100 & 5 & 40 & 40 & 10  & $F/N_{bi}$ \\
			1 & 5 & 40 & 40 & 25  & $N_{mono}$  & 100 & 5 & 40 & 40 & 25  & $N_{mono}$ \\
			1 & 5 & 40 & 40 & 50  & $N_{mono}$  & 100 & 5 & 40 & 40 & 50  & $N_{mono}$ \\
			1 & 5 & 40 & 40 & 100 & $N_{mono}$  & 100 & 5 & 40 & 40 & 100 & $N_{bi}$ \\
			1 & 5 & 64 & 64 & 10  & $N_{bi}$    & 100 & 5 & 64 & 64 & 10  & $K$ \\
			1 & 5 & 64 & 64 & 25  & $N_{bi}$    & 100 & 5 & 64 & 64 & 25  & $N_{mono}$ \\
			1 & 5 & 64 & 64 & 50  & $N_{bi}$    & 100 & 5 & 64 & 64 & 50  & $N_{bi}$ \\
			1 & 5 & 64 & 64 & 100 & $N_{bi}$    & 100 & 5 & 64 & 64 & 100 & $N_{bi}$ \\
			\hline
			\hline
			1 & 10 & 16 & 16 & 10  & $F/N_{bi}$ & 100 & 10 & 16 & 16 & 10  & $F$ \\
			1 & 10 & 16 & 16 & 25  & $N_{bi}$   & 100 & 10 & 16 & 16 & 25  & $N_{bi}$ \\
			1 & 10 & 16 & 16 & 50  & $N_{mono}$ & 100 & 10 & 16 & 16 & 50  & $F$ \\
			1 & 10 & 16 & 16 & 100 & $N_{mono}$ & 100 & 10 & 16 & 16 & 100 & $K$ \\
			1 & 10 & 40 & 40 & 10  & $F/N_{bi}$ & 100 & 10 & 40 & 40 & 10  & $F$ \\
			1 & 10 & 40 & 40 & 25  & $N_{bi}$   & 100 & 10 & 40 & 40 & 25  & $N_{bi}$ \\
			1 & 10 & 40 & 40 & 50  & $N_{mono}$ & 100 & 10 & 40 & 40 & 50  & $N_{mono}$ \\
			1 & 10 & 40 & 40 & 100 & $N_{bi}$ 	& 100 & 10 & 40 & 40 & 100 & $N_{mono}$ \\
			1 & 10 & 64 & 64 & 10  & $N_{bi}$   & 100 & 10 & 64 & 40 & 10  & $F$ \\
			1 & 10 & 64 & 64 & 25  & $N_{bi}$   & 100 & 10 & 64 & 40 & 25  & $N_{bi}$ \\
			1 & 10 & 64 & 64 & 50  & $N_{bi}$   & 100 & 10 & 64 & 40 & 50  & $N_{mono}$ \\
			1 & 10 & 64 & 64 & 100 & $N_{bi}$   & 100 & 10 & 64 & 40 & 100 & $N_{bi}$ \\
			\hline	
			\hline
			1 & 25 & 16 & 16 & 10  & $F/N_{bi}$ & 100 & 25 & 16 & 16 & 10  & $F/N_{mono}$ \\
			1 & 25 & 16 & 16 & 25  & $N_{mono}$ & 100 & 25 & 16 & 16 & 25  & $N_{mono}$ \\
			1 & 25 & 16 & 16 & 50  & $N_{mono}$ & 100 & 25 & 16 & 16 & 50  & $N_{mono}$ \\
			1 & 25 & 16 & 16 & 100 & $N_{mono}$ & 100 & 25 & 16 & 16 & 100 & $N_{mono}$ \\
			1 & 25 & 40 & 40 & 10  & $F/N_{bi}$ & 100 & 25 & 40 & 40 & 10  & $F/N_{mono}$ \\
			1 & 25 & 40 & 40 & 25  & $N_{mono}$ & 100 & 25 & 40 & 40 & 25  & $N_{mono}$ \\
			1 & 25 & 40 & 40 & 50  & $N_{mono}$ & 100 & 25 & 40 & 40 & 50  & $N_{mono}$ \\
			1 & 25 & 40 & 40 & 100 & $N_{mono}$ & 100 & 25 & 40 & 40 & 100 & $N_{mono}$ \\
			1 & 25 & 64 & 64 & 10  & $F$        & 100 & 25 & 64 & 64 & 10  & $K$ \\
			1 & 25 & 64 & 64 & 25  & $N_{bi}$   & 100 & 25 & 64 & 64 & 25  & $N_{mono}$ \\
			1 & 25 & 64 & 64 & 50  & $N_{mono}$ & 100 & 25 & 64 & 64 & 50  & $N_{bi}$ \\
			1 & 25 & 64 & 64 & 100 & $N_{bi}$   & 100 & 25 & 64 & 64 & 100 & $F$ \\
			\hline
		\end{tabular}
	}
\end{table*}

\begin{table*}[]
	\centering
	\caption{Parameters for flexocytes of puller type shown in Fig.~2 (d). For all the systems $\tilde{\kappa}=2$ and $\tilde{k}_{\rm S}=1000$.}\label{tab:pull}
	\rowcolors{2}{gray!25}{white}
	{\footnotesize
		\begin{tabular}{p{1cm}p{1cm}p{1cm}p{1cm}p{1cm}p{2cm}||p{1cm}p{1cm}p{1cm}p{1cm}p{1cm}p{2cm}}
			\hline
			$\tilde{k}_{\rm A}$ & $\tilde{\gamma}_{\rm s}$ & $N_\textrm{r,\textrm{push}}$ & $N_\textrm{r,\textrm{pull}}$ & Pe & Flexocyte type & $\tilde{k}_{\rm A}$ & $\tilde{\gamma}_{\rm s}$ & $N_\textrm{r,\textrm{push}}$ & $N_\textrm{r,\textrm{pull}}$ & Pe & Flexocyte type \\
			\hline
			\hline
			1 & 1 & - & 16 & 10  & $F$   & 100 & 1 & - & 16 & 10  & $F$ \\
			1 & 1 & - & 16 & 25  & $F/K$ & 100 & 1 & - & 16 & 25  & $K$ \\
			1 & 1 & - & 16 & 50  & $K$   & 100 & 1 & - & 16 & 50  & $K$ \\
			1 & 1 & - & 16 & 100 & $K$   & 100 & 1 & - & 16 & 100 & $K$ \\
			1 & 1 & - & 40 & 10  & $F$   & 100 & 1 & - & 40 & 10  & $K$ \\
			1 & 1 & - & 40 & 25  & $K$   & 100 & 1 & - & 40 & 25  & $K$ \\
			1 & 1 & - & 40 & 50  & $K$   & 100 & 1 & - & 40 & 50  & $K$ \\
			1 & 1 & - & 40 & 100 & $K$   & 100 & 1 & - & 40 & 100 & $K$ \\
			1 & 1 & - & 64 & 10  & $F$   & 100 & 1 & - & 64 & 10  & $F$ \\
			1 & 1 & - & 64 & 25  & $K$   & 100 & 1 & - & 64 & 25  & $K$ \\
			1 & 1 & - & 64 & 50  & $K$   & 100 & 1 & - & 64 & 50  & $K$ \\
			1 & 1 & - & 64 & 100 & $K$   & 100 & 1 & - & 64 & 100 & $K$ \\
			\hline	
			\hline	
			1 & 5 & - & 16 & 10  & $F$   & 100 & 5 & - & 16 & 10  & $F$ \\
			1 & 5 & - & 16 & 25  & $F$   & 100 & 5 & - & 16 & 25  & $F$ \\
			1 & 5 & - & 16 & 50  & $F/K$ & 100 & 5 & - & 16 & 50  & $K$ \\
			1 & 5 & - & 16 & 100 & $K$   & 100 & 5 & - & 16 & 100 & $K$ \\
			1 & 5 & - & 40 & 10  & $F$   & 100 & 5 & - & 40 & 10  & $F$ \\
			1 & 5 & - & 40 & 25  & $F$   & 100 & 5 & - & 40 & 25  & $F$ \\
			1 & 5 & - & 40 & 50  & $K$   & 100 & 5 & - & 40 & 50  & $K$ \\
			1 & 5 & - & 40 & 100 & $N$   & 100 & 5 & - & 40 & 100 & $K$ \\
			1 & 5 & - & 64 & 10  & $F$   & 100 & 5 & - & 64 & 10  & $F$ \\
			1 & 5 & - & 64 & 25  & $F$   & 100 & 5 & - & 64 & 25  & $F$ \\
			1 & 5 & - & 64 & 50  & $N$   & 100 & 5 & - & 64 & 50  & $K$ \\
			1 & 5 & - & 64 & 100 & $N$   & 100 & 5 & - & 64 & 100 & $K$ \\
			\hline
			\hline
			1 & 10 & - & 16 & 10  & $F$ & 100 & 10 & - & 16 & 10  & $F$ \\
			1 & 10 & - & 16 & 25  & $F$ & 100 & 10 & - & 16 & 25  & $F$ \\
			1 & 10 & - & 16 & 50  & $F$ & 100 & 10 & - & 16 & 50  & $F$ \\
			1 & 10 & - & 16 & 100 & $F$ & 100 & 10 & - & 16 & 100 & $K$ \\
			1 & 10 & - & 40 & 10  & $F$ & 100 & 10 & - & 40 & 10  & $F$ \\
			1 & 10 & - & 40 & 25  & $F$ & 100 & 10 & - & 40 & 25  & $F$ \\
			1 & 10 & - & 40 & 50  & $F$ & 100 & 10 & - & 40 & 50  & $F$ \\
			1 & 10 & - & 40 & 100 & $N$ & 100 & 10 & - & 40 & 100 & $K$ \\
			1 & 10 & - & 64 & 10  & $F$ & 100 & 10 & - & 40 & 10  & $F$ \\
			1 & 10 & - & 64 & 25  & $F$ & 100 & 10 & - & 40 & 25  & $F$ \\
			1 & 10 & - & 64 & 50  & $N$ & 100 & 10 & - & 40 & 50  & $N$ \\
			1 & 10 & - & 64 & 100 & $N$ & 100 & 10 & - & 40 & 100 & $K$ \\
			\hline	
			\hline
			1 & 25 & - & 16 & 10  & $F$ & 100 & 25 & - & 16 & 10  & $F$ \\
			1 & 25 & - & 16 & 25  & $F$ & 100 & 25 & - & 16 & 25  & $F$ \\
			1 & 25 & - & 16 & 50  & $F$ & 100 & 25 & - & 16 & 50  & $F$ \\
			1 & 25 & - & 16 & 100 & $F$ & 100 & 25 & - & 16 & 100 & $F$ \\
			1 & 25 & - & 40 & 10  & $F$ & 100 & 25 & - & 40 & 10  & $F$ \\
			1 & 25 & - & 40 & 25  & $F$ & 100 & 25 & - & 40 & 25  & $F$ \\
			1 & 25 & - & 40 & 50  & $F$ & 100 & 25 & - & 40 & 50  & $F$ \\
			1 & 25 & - & 40 & 100 & $F$ & 100 & 25 & - & 40 & 100 & $F$ \\
			1 & 25 & - & 64 & 10  & $F$ & 100 & 25 & - & 64 & 10  & $F$ \\
			1 & 25 & - & 64 & 25  & $F$ & 100 & 25 & - & 64 & 25  & $F$ \\
			1 & 25 & - & 64 & 50  & $F$ & 100 & 25 & - & 64 & 50  & $F$ \\
			1 & 25 & - & 64 & 100 & $F$ & 100 & 25 & - & 64 & 100 & $F$ \\
			\hline
		\end{tabular}
	}
\end{table*}

\bibliography{biblio}